\documentclass[a4paper,10pt,superscriptaddress,showkeys,aps,pra,showpacs,twocolumn]{revtex4-2}
\usepackage{amssymb, amsmath, amsfonts,bm} 
\usepackage{siunitx} 
\usepackage{IEEEtrantools} 
\usepackage{graphics,xcolor} 
\usepackage{epsfig}
\usepackage{pst-all}
\usepackage{hyperref} 
\usepackage[mathlines,running,modulo]{lineno}
\usepackage{relsize}
\usepackage[utf8]{inputenc}
\usepackage[toc]{appendix}
\usepackage{filecontents}
\usepackage{latexsym} 
\usepackage[version=3]{mhchem} 
\usepackage{xcolor}

\begin{document}

\title{Detecting quantum spin liquid on Kitaev model through a superconducting junction}
\author{A. R. Moura}
\email{antoniormoura@ufv.br}
\affiliation{Departamento de Física, Universidade Federal de Viçosa, 36570-900, Viçosa, Minas Gerais, Brazil}
\author{L. V. Santos}
\date{\today}

\begin{abstract}
The Kitaev model belongs to an unconventional class of two-dimensional spin systems characterized by anisotropic, bond-dependent 
interactions that give rise to Quantum Spin Liquid (QSL) states. These exotic phases, marked by the absence of magnetic ordering even at zero 
temperature, support fractionalized excitations and emergent gauge fields. A particularly compelling feature of the Kitaev model is its 
exact solvability, which reveals low-energy excitations in the form of itinerant Majorana fermions-quasiparticles that obey non-Abelian 
statistics and are of central interest in topological quantum computation due to their inherent robustness against local perturbations and
decoherence. Despite extensive theoretical advancements, the experimental identification of QSLs remains challenging, as conventional 
magnetic probes fail to detect their defining properties. In this work, we present a theoretical investigation of spin current injection 
from a superconducting metal into a Kitaev quantum spin liquid. By employing a spintronic framework, we derive the dynamics of the injected 
spin current and demonstrate how its signatures can be traced back to the underlying Majorana excitations in the spin liquid phase. 
Superconductivity plays a pivotal role in this context, not only as a source of coherent quasiparticles but also as a platform with potential 
for interfacing with topological quantum devices. The interplay between superconductivity and the Kitaev QSL enables a novel mechanism for
probing the spin transport mediated by Majorana fermions. Our analysis contrasts the Kitaev-superconductor interface with conventional 
ferromagnetic junctions, where spin transport is carried by magnons, and highlights distinctive features in the spin current response. 
These findings open new directions for the detection of QSLs and contribute to the broader effort of integrating topological quantum 
materials into scalable quantum technologies.

\end{abstract}


\maketitle

\section{Introduction and motivation}
\label{sec.introduction}

In recent years, the study of spin-dependent transport phenomena has emerged as a pivotal avenue in the field of spintronics, 
motivated by the need to harness spin currents in diverse materials beyond conventional ordered magnetic models. For instance, 
spin current driven by magnons has been observed in ferromagnetic (FM) \cite{prl88.117601,prb83.144402,jpcs200.062030}, 
antiferromagnetic (AF) \cite{rmp90.015005,prb89.140406,prb90.094408,prb93.054412,prb94.014412,prl116.186601},
 and paramagnetic (PM) insulators \cite{prl116.186601,naturecomm10.1,naturephys13.987}. Notably, materials lacking 
long-range magnetic order, such as Quantum Spin Liquids (QSLs), have demonstrated a surprising capacity to 
sustain and mediate spin transport, thereby challenging traditional paradigms of magnetic order as a prerequisite for spin current 
propagation. As is common in spintronic experiments, various proposals have been put forward to utilize junctions composed 
of two or more layers of distinct materials in order to investigate the characteristics of non-conventional spin currents. 
In particular, S. Chatterjee and S. Sachdev proposed that the spin current–voltage relationship could serve as a tool to identify 
different types of QSL excitations \cite{prb92.165113}. Analogous strategies have been employed to induce excitations and to identify characteristic 
signatures of spin transport in the more general SU(2)-symmetric Kitaev model \cite{prb98.155105,prb104.l060403}.
An alternative strategy for detecting spinons in Mott insulators is presented in Ref.\cite{prb88.041405}, while Ref.\cite{prr2.033439} 
proposes a method for identifying QSL based on the analysis of voltage noise spectra in junction devices.

Among the most prominent theoretical frameworks supporting QSL phases is the Kitaev model, a two-dimensional spin-$1/2$ system
characterized by direction-dependent interactions on a honeycomb lattice \cite{ap321.2,prb97.115142,knolle}. 
This model supports exotic excitations in the form of itinerant 
Majorana fermions coupled to a static $Z_2$ gauge field, rendering it an ideal candidate for exploring topologically nontrivial spin transport. 
Although originally introduced as a highly idealized and theoretical construct, the Kitaev model has since motivated considerable experimental 
efforts aimed at its physical realization. In recent years, numerous materials have been proposed as promising platforms for hosting 
two-dimensional QSL phases consistent with Kitaev-type interactions \cite{science367.0668,nrp1.264,pr950.1}. Nevertheless, many of these 
candidate systems exhibit additional interactions, beyond those prescribed by the pure Kitaev model, that tend to stabilize magnetically 
ordered ground states at sufficiently low temperatures. Among the proposed materials, the ruthenium-based compound $\alpha$-\ce{RuCl_3} 
has emerged as one of the most compelling candidates for realizing Kitaev physics in spin-$1/2$ 
systems \cite{prl114.147201,nphys13.1079,science356.1055,npj3.1,aplm10.080903}.

As one can imagine, the electrical insulation intrinsic to QSLs imposes significant experimental challenges, redirecting attention to indirect detection 
methods such as thermal and spin current measurements. For instance, in the \ce{Sr_2CuO_3} spin chain, the spin current is driven by
spinons, which have been verified by the spin Seebeck effect \cite{naturephys13.30}. In particular, recent proposals have focused 
on spin injection from adjacent normal metals (NM) into Kitaev materials, where spin-flip processes at the interface can excite Majorana 
modes and generate distinctive spin transport signatures \cite{jmmm624.173024}. These processes are strongly influenced by interface 
interactions, notably the spin-transfer torque mechanism, which facilitates angular momentum exchange without requiring external magnetic fields.

Although the majority of interface studies focus on normal-state transport, the injection of charge currents at 
superconducting (SC) interfaces has also been extensively investigated over the past decades. In particular, 
the presence of spin-polarized quasiparticles in $s$-wave superconductors has been attributed to the injection of spin-polarized 
charge currents, accompanied by spin accumulation and spin diffusion within the superconducting medium \cite{PRL55.1790,PRB37.5326,APL65.1460}.
On a parallel front, superconducting systems have also been identified as efficient mediators of spin current, particularly when 
interfaced with magnetic insulators. For example, in Ref. \cite{prl100.047002}, the impact of superconductivity on spin transport was 
investigated through measurements of the Gilbert damping in \ce{Ni_{80}Fe_{20}} films grown on \ce{Nb}, while the spin dynamics at interfaces 
comprising superconducting \ce{NbN} films and the ferromagnetic insulator \ce{GdN} have been investigated in Ref. \cite{prb97.224414}.
Theoretical studies describing spin current injection at superconductor/ferromagnetic insulating interfaces have been developed in Refs. \cite{prb93.064421,prb96.024414,prb99.144411,jmmm494.165813,prb102.024412}. Notably, even paramagnetic insulators, long considered 
spin-inactive due to the absence of magnetic order, exhibit measurable spin transport when coupled to superconductors, further 
highlighting the generality of spin current phenomena across ordered and disordered quantum phases. 

In this context, the interface between QSLs described by the Kitaev model and superconducting systems provides a fertile platform 
for investigating unconventional spin-transport mechanisms \cite{prr4.023251}. Understanding how spin currents are generated, transmitted, 
and detected in such hybrid architectures is essential not only for probing the elusive properties of QSLs, but also for advancing 
spintronic concepts that exploit topological and quantum-coherent effects. In this work, we develop a microscopic theoretical framework to 
describe spin-flip scattering processes of superconducting quasiparticles at the interface, which enables spin current injection into 
the Kitaev QSL. Building upon this framework, we employ linear-response theory to calculate the spin current generated under weak 
nonequilibrium conditions, which are induced by a chemical potential imbalance and constrained by the conditions imposed by the 
Kitaev and SC layers. Our analysis thus connects these domains by establishing the theoretical foundations and physical consequences 
of spin current injection at the interface between Kitaev-type QSLs and superconducting materials. 
In this scenario, quasiparticles originating from the superconducting side interact with Majorana fermions at the interface 
via spin-flip scattering processes, thereby transferring angular momentum into the Kitaev system through a mechanism analogous 
to spin-transfer torque (STT) \cite{jmmm159.l1,prb54.9353}. As we demonstrate, the resulting spin transport reflects the propagation of 
spin information mediated by Majorana fermions in a spin-liquid state, leading to qualitative differences from the conventional case of 
ordered magnetic insulators with magnonic excitations and offering a distinctive signature of the Kitaev QSL phase.It is important 
to note that, although the primary objective is the investigation of the Kitaev spin liquid, the resulting spin current is sufficiently 
general to be applicable to other realizations of QSL states, as well as to certain magnetically ordered phases.

{The structure of the paper is organized as follows. In Sec. (\ref{sec.model}), we describe the model under consideration. In Sec. (\ref{sec.model}), 
we provide a review of the BCS formalism and the Kitaev model. The formulation of the injected spin current is presented in Sec. (\ref{sec.spincurrent}), 
while further details concerning the Kitaev spin–spin correlations are discussed in Sec. (\ref{sec.kitaev_ssc})Finally, the principal results concerning
the spin current and spin conductance, as well as the concluding remarks, are presented in Sec. (\ref{sec.results}) and Sec. (\ref{sec.conlusion}), respectively.

\section{Model description}
\label{sec.model}
The complete Hamiltonian is written as $H=H_{SC}+H_K+H_{sd}$, where $H_{SC}+H_K$ describes the free Hamiltonian, which exhibits an exact
solution, while $H_{sd}$ represents the interface interaction term.  The SC term is given by
\begin{equation}
    \label{eq.Hsc}
    H_{SC}=\sum_{k\sigma}\epsilon_k c_{k\sigma}^\dagger c_{k\sigma}-g\sum_{kk^\prime}c_{k\uparrow}^\dagger c_{-k\downarrow}^\dagger c_{-k^\prime\downarrow}c_{k^\prime\uparrow},
\end{equation}
where $\epsilon_k = \hbar^2 k^2 / 2m$ denotes the kinetic energy, $c_{k\sigma}$ ($c_{k\sigma}^\dagger$) are the 
electron annihilation (creation) operators, and $g$ is the effective superconducting coupling constant. 
The summation over electronic momentum is restricted to a thin shell of width $2\hbar\omega_D$ around the Fermi surface,
$|\epsilon_k - \epsilon_F| \leq \hbar\omega_D$, where $\hbar\omega_D$ and $\epsilon_F$ represent the Debye and Fermi energies, 
respectively. Typical energy scales are $\epsilon_F \sim 10$ eV and $\hbar\omega_D \sim 10^{-1}$ eV. A detailed description of
the SC Hamiltonian is developed in Sec. \ref{sec.sc}.

\begin{figure}[h]
\centering \epsfig{file=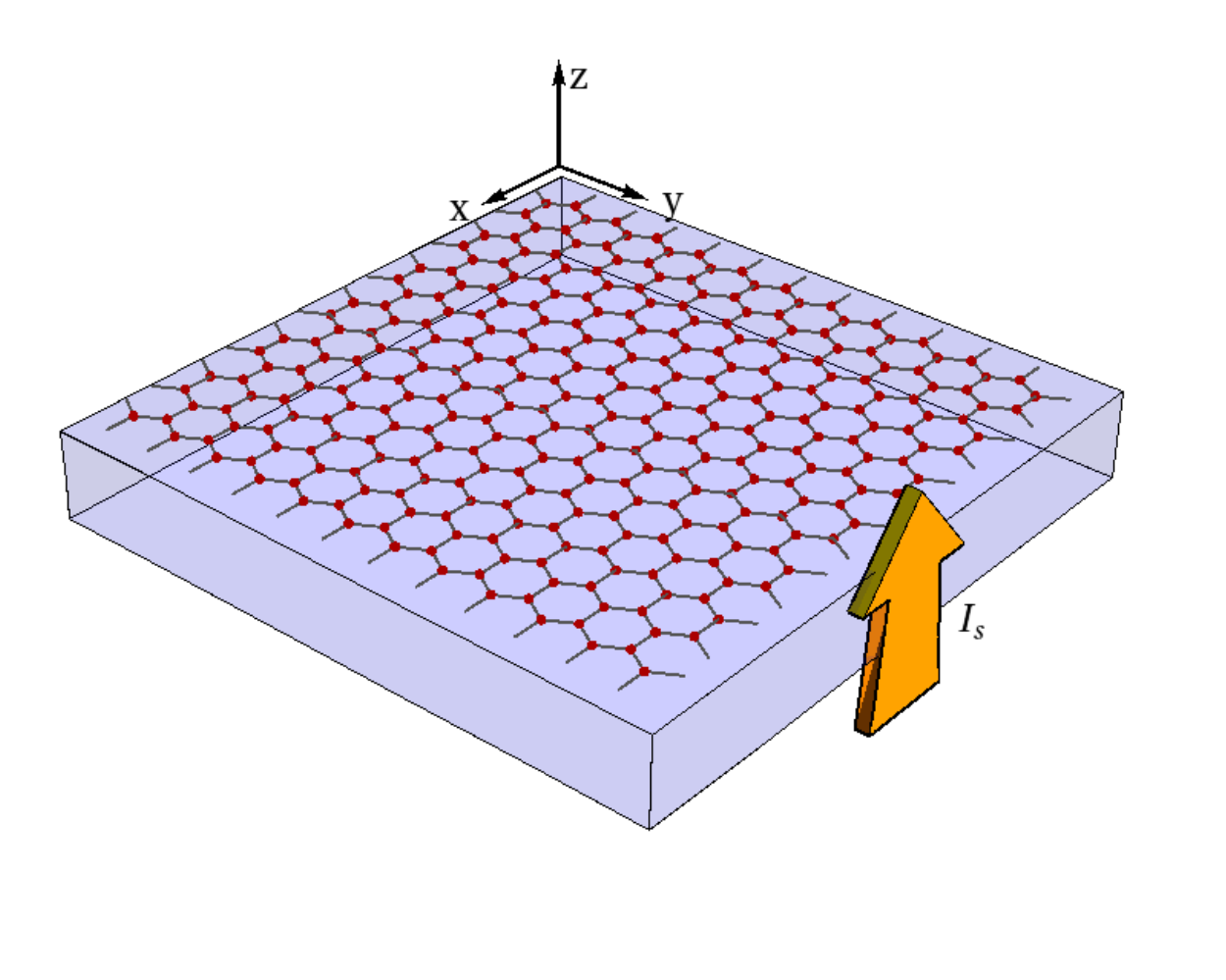,width=0.9\linewidth}
\caption{A Kitaev layer deposited on a SC substrate sustains an interfacial spin current $I_s$ driven by a 
chemical potential imbalance between spin-up and spin-down electrons.}
\label{fig.sample}
\end{figure}

The Kitaev model is defined as a bond-directional Ising model comprising spin-$1/2$ particles located at the vertices of a honeycomb 
lattice. The corresponding Hamiltonian reads
\begin{equation}
    \label{eq.HK}
    H_K=-J_x\sum_{\langle ij\rangle_x} \sigma_i^x \sigma_j^x - J_y\sum_{\langle ij\rangle_y} \sigma_i^y \sigma_j^y -
    J_z\sum_{\langle ij\rangle_z} \sigma_i^z \sigma_j^z,
\end{equation}
where $\langle ij\rangle_a$ denotes a link along the $a$-direction [see Fig.\ref{fig.honeycomb}], and the $\sigma^a$ operators represent 
Pauli matrices. The anisotropic exchange constants satisfy the constraint $J_x + J_y + J_z = 1$, defining qualitatively distinct 
phases depending on their relative magnitudes. The gapped A phase arises when $J_x > J_y + J_z$ (and cyclic permutations thereof), 
whereas the gapless B phase emerges for $J_x < J_y + J_z$. In this work, we consider the isotropic limit with $J_x = J_y = J_z = J > 0$. 
For the material $\alpha$-\ce{RuCl_3}, which closely approximates the isotropic Kitaev model, the coupling constant is estimated 
to be $J \approx 8$ meV\cite{prl114.147201,natmat15.733}. The development of the Kitaev model and its intrinsic properties is 
elaborated in Sec. \ref{sec.kitaev}.

The interaction at the interface is described by the $sd$-exchange Hamiltonian,
\begin{equation}
H_{sd} = -J_{sd} \sum_i \bm{s}_i \cdot \bm{\sigma}_i,
\end{equation}
where $J_{sd}>0$ denotes the coupling strength between the spins of conduction electrons and localized magnetic moments at the interface. 
The operator $\bm{s}_i$ represents the spin of conduction electrons in the SC, and $\bm{\sigma}_i$ denotes the localized spin operator in the 
FM layer. The SC region is defined as the half-space $z<0$, while the FM monolayer is situated in the $z=0$ plane (just above the $xy$-plane).
A graphical representation is shown in Fig. (\ref{fig.sample}).

The longitudinal interaction term $s_i^z \sigma_i^z$ corresponds to processes conserving spin projection and, therefore, does not contribute to 
the net transfer of angular momentum across the interface. Instead, spin current injection is driven by processes involving a change in the 
magnetization component along the $z$-axis, the direction of spin current polarization. Such processes are mediated by the transverse spin 
components, specifically the ladder operators $\sigma_i^\pm = \sigma_i^x \pm i \sigma_i^y$.

Spin-flip scattering, which facilitates angular momentum transfer, is closely related to the concept of spin-mixing conductance. 
This quantity, emerging naturally in the Landau-Lifshitz-Gilbert formalism, quantifies the efficiency with which spin angular momentum is 
transmitted through the interface \cite{prl88.117601,prb66.224403}. Accordingly, in terms of electronic operators, the effective 
$sd$ Hamiltonian can be written as
\begin{equation}
H_{sd}=-J_{sd} \sum_i \left( \sigma_i^+ c_{i\downarrow}^\dagger c_{i\uparrow} + \sigma_i^- c_{i\uparrow}^\dagger c_{i\downarrow} \right),
\end{equation}
where $c_{i\sigma}^\dagger$ ($c_{i\sigma}$) creates (annihilates) a conduction electron with spin $\sigma=\pm 1/2$ at site $i$.
Subsequently, we should replace the electronic operators with the quasiparticle operators of the superconducting Hamiltonian.

Due to the local nature and spatial variability of the exchange interaction, it is challenging to assign a uniform value for 
$J_{sd}$ across the entire interface. Based on his analysis of resistivity in magnetic metal alloys, Kondo estimated that $J_{sd}$ is 
typically on the order of a few percent of the Fermi energy \cite{ptp32.37}. In the present context, we adopt a phenomenological value for 
$J_{sd}$ that is taken to be on the order of the characteristic magnetic exchange energy scale between neighboring magnetic sites.
To proceed analytically, we perform a Fourier transform of the electron operators:
\begin{equation}
c_{i\sigma} = \sqrt{\frac{2}{N_e}} \sum_k c_{k\sigma} e^{i \bm{k} \cdot \bm{r}_i},
\end{equation}
where $N_e = \rho_e V_e$ is the total number of electrons, with $V_e = L_x L_y L_z$ being the volume of the SC region and $\rho_e$ 
its electronic density. The SC domain is confined to $-L_z \leq z\leq 0$, and the wavenumbers are quantized as 
$k_a=2\pi n_a/L_a$, with $n_x, n_y \in \mathbb{Z}$ and $n_z \in \mathbb{N}$. This formulation serves as the foundation for subsequent analysis of spin transport across the SC/FM interface.

\section{Free Hamiltonians}

\subsection{Superconducting Hamiltonian}
\label{sec.sc}
To enable spin current injection from the superconductor, a population imbalance between spin-up and spin-down electrons is 
required. Incorporating the chemical potentials $\mu_\uparrow$ and $\mu_\downarrow$ for spin-up and spin-down electrons, respectively, 
we replace the Hamiltonian $H_{SC}$ by the grand-canonical Hamiltonian $K_{SC}=H_{SC}-\sum_\sigma \mu_\sigma N_\sigma$, 
where $N_\sigma=\sum_k c_{k\sigma}^\dagger c_{k\sigma}$. Using the spinor notation, $K_{SC}$ reads
\begin{align}
&K_\textrm{SC}=\sum_k\Psi_k^\dagger\begin{pmatrix} \xi_k-\Delta\mu/2 & -\Delta \\
-\bar{\Delta} &-\xi_k-\Delta\mu/2\end{pmatrix}\Psi_k+\nonumber\\
&+\frac{|\Delta|^2}{g} + \sum_k \left(\xi_k + \frac{\Delta\mu}{2}\right),
\end{align}
where $\Psi_k^\dagger=(c_{k\uparrow}^\dagger\ c_{-k\downarrow})$, and $\Delta\mu=\mu_\uparrow-\mu_\downarrow$ acts as an 
effective magnetic field that lifts the spin degeneracy. The quartic interaction term has been decoupled via a mean-field 
approximation by introducing the superconducting gap parameter $\Delta = g \sum_k \langle c_{-k\downarrow} c_{k\uparrow} \rangle$. 
In this expression, the electronic energy $\xi_k = \epsilon_k - \mu_m$ is defined relative to the mean chemical potential,
$\mu_m = (\mu_\uparrow + \mu_\downarrow)/2$. The BCS Hamiltonian can be diagonalized by 
introducing new fermionic operators through the Bogoliubov transformation $b_{k\uparrow}=\bar{u}_k c_{k\uparrow} + v_k c_{-k\downarrow}^\dagger$
and $b_{k\downarrow}=\bar{u}_k c_{k\downarrow} - v_k c_{-k\uparrow}^\dagger$,
where the coherence factors are defined as $u_k = e^{-i\varphi/2} \cos\theta_k$,
and $v_k = e^{i\varphi/2}\sin\theta_k$. Here, $\varphi$ denotes the phase of the superconducting gap, such that 
$\Delta = e^{i\varphi}|\Delta|$, and, in order to eliminate the off-diagonal terms, we define the $\theta_k$ angle
through the relation $\tan2\theta_k=|\Delta|/\xi_k$. The resulting diagonalized BCS Hamiltonian is given by
\begin{equation}
K_\textrm{SC}=K_0+\sum_k \left( E_{k\uparrow} b_{k\uparrow}^\dagger b_{k\uparrow} + E_{k\downarrow} b_{k\downarrow}^\dagger b_{k\downarrow} \right),
\end{equation}
where $K_0 = |\Delta|^2 / g + \sum_k (\xi_k - E_k)$ is the ground-state energy, and $E_{k\sigma}=E_k-\sigma \Delta\mu/2$ 
accounts for the spin-dependent quasiparticle energies. Note that, while the superconducting ground state consists of Cooper pairs, 
the elementary excitations are quasiparticles, commonly referred to as Bogoliubov quasiparticles, with energy dispersion given by 
$E_k= \sqrt{\xi_k^2 + |\Delta|^2}$. It is straightforward to verify that $K_0$ is less than the normal 
ground state energy, and the electron pairing provides a stable SC state below the critical temperature. 
From this Hamiltonian, the temperature dependence of the gap equation is obtained from the self-consistent
equation.
\begin{equation}
\label{eq.gapsc}
\Delta(T) = \sum_{k,\sigma} \frac{|g|^2 \Delta(T)}{4E_k} \tanh\left( \frac{E_{k\sigma}}{2k_B T} \right),
\end{equation}
which determines the temperature dependence of the superconducting gap. As usual, the superconducting critical 
temperature is defined as the point at which the gap vanishes. To accurately evaluate the temperature dependence of the injected spin current, 
the superconducting gap is obtained as the solution of the above equation for a given chemical imbalance \(\Delta \mu\).

Due to the structure of the quasiparticle spectrum, a finite difference in chemical potential between spin-up and spin-down 
states enhances processes involving the annihilation (creation) of spin-up (spin-down) quasiparticles. Consequently, a positive value 
of $\Delta\mu$ induces a net spin current from the superconductor into the adjacent magnetic insulator.  For $\Delta\mu=0$, the model exhibits 
the usual continuous transition temperature at $T_{c0}=T(\Delta\mu=0)$, presenting a superconducting phase for $T<T_{c0}$, and the maximum gap 
$|\Delta_0|$ is achieved at zero temperature. However, polarizing effects such as external magnetic fields and chemical potential imbalance $\Delta\mu$ tend to 
destabilize the superconducting phase, reducing the critical temperature to an effective value $T_c<T_{c0}$ \cite{jpcs24.1029,jlpt18.297}. 
The gap decreases monotonically with increasing chemical imbalance, providing different regimes depending on the $\Delta\mu$ value. 
When $\Delta\mu \leq 1.20|\Delta_0|$, the superconducting transition is continuous and, for $\Delta\mu=1.20|\Delta_0|$, the 
reduced critical temperature is given by $T_c\approx 0.59\ T_{c0}$. In the intermediate regime $1.20|\Delta_0|<\Delta\mu<1.42|\Delta_0|$, 
the gap exhibits a discontinuous jump at $T_c$. Beyond $\Delta\mu > 1.42|\Delta_0|$, superconductivity is destroyed even at zero 
temperature. In this work, we restrict our analysis to the continuous regime, for which $\Delta\mu_\textrm{max}=1.20\Delta_0|$. 
Additionally, for $T \lesssim T_c$, the gap $|\Delta(T)|$ is of order $0.1|\Delta_0|$.
As an illustration, we take $T_c(\Delta\mu = 0) \approx 50\,\mathrm{K}$, which corresponds to an energy gap $|\Delta_0| = 7.62\,\mathrm{meV}$. The maximum
chemical imbalance is then $\Delta\mu_{\mathrm{max}} = 9.16\,\mathrm{meV}$, leading to a reduced transition temperature $T_c \approx 30\,\mathrm{K}$.

\subsection{Kitaev Hamiltonian}
\label{sec.kitaev}
We now present a concise overview of the principal characteristics of the Kitaev model, employing the $SO(4)$ representation.
Our focus lies on the solution within the flux-free sector, which is the relevant case for the present study. A comprehensive review 
of the Kitaev model and its various representations is available in Ref.~\cite{prb97.115142}.

We focus on the model in the absence of an external magnetic field. The inclusion of a Zeeman term modifies the dispersion relation, 
opening a gap in the energy spectrum. Without a magnetic field, the system supports gapless relativistic excitations, which 
simplifies the analytical treatment of spin currents. Nevertheless, numerical methods can extend the analysis to include magnetic 
fields, thereby improving the agreement between theoretical predictions and experimental observations of field-induced QSL phases 
in $\alpha$-\ce{RuCl_3}~\cite{prb95.180411,prl119.037201,prl119.227208,prl120.117204}.

The exact solvability of the Kitaev model arises from the existence of an extensive number of local conserved quantities, allowing the 
decomposition of the Hilbert space into distinct topological sectors. One such conserved quantity is the plaquette operator (or flux operator), 
defined on a hexagonal loop as $W_p = \sigma_1^x \sigma_2^y \sigma_3^z \sigma_4^x \sigma_5^y \sigma_6^z$, which commutes with both the Hamiltonian 
and other plaquette operators. As $W_p^2 = I$, its eigenvalues are $\pm 1$, enabling a classification of states into flux sectors. 
The ground state resides in the flux-free sector, where $W_p = 1$ for all plaquettes. States with $W_p = -1$ correspond to vortex 
excitations, which always occur in pairs and require a two-flux excitation energy of approximately $\Delta_F \approx 0.26 J=2.08$ meV in the isotropic case.
As will be shown, the Kitaev gap imposes a lower bound on the chemical imbalance, and by additionally considering the upper limit of the superconducting 
phase, we obtain the restriction $\Delta_F \leq \Delta\mu \leq \Delta\mu_\textrm{max}$. Note that the model is characterized by two distinct energy scales. The lower 
scale is determined by the Kitaev gap, whereas the upper scale is specified by the maximum chemical potential that still supports a continuous SC phase transition. 
Assuming $T_{c0} = 50\,\mathrm{K}$, which results in $\Delta\mu_\textrm{max}=9.16\,\mathrm{meV}$, and $J = 8\,\mathrm{meV}$, we obtain a correspondence between these 
energy scales such that $\Delta\mu_\textrm{max} \approx 4.40\,\Delta_F$.

To diagonalize the Hamiltonian, the spin operators are mapped to Majorana fermions (MFs) via
\begin{equation}
\sigma_i^a = i c_i b_i^a,
\end{equation}
where $c_i^\dagger = c_i$ and $(b_i^a)^\dagger = b_i^a$ for $a = x, y, z$. These MFs satisfy the canonical anticommutation 
relations $\{b_i^a, b_j^b\} = 2\delta_{ij}\delta_{ab}$, $\{c_i, c_j\} = 2\delta_{ij}$, and $\{b_i^a, c_j\} = 0$. This fermionic representation 
faithfully reproduces the spin algebra. However, while the original Hilbert space has dimension $2^N$ for $N$ sites, the MF 
representation spans a larger space of dimension $2^{2N}$, containing unphysical states. To restrict to the physical subspace, we use 
the identity $I = D_i = -i\sigma_i^x\sigma_i^y\sigma_i^z = b_i^x b_i^y b_i^z c_i$ and define the projection operator $P = \prod_i(1 + D_i)/2$, 
such that $|\Psi_\textrm{phys}\rangle = P|\Psi\rangle$. Since $[H, D_i] = 0$ and $[W_p, D_i] = 0$, the physical subspace remains closed 
under both the Hamiltonian and flux operators.

The Kitaev Hamiltonian expressed in terms of MFs becomes
\begin{equation}
H_K = i \sum_{\langle ij\rangle_a} J_a u_{\langle ij\rangle_a} c_i c_j,
\end{equation}
with the bond operator defined as $u_{\langle ij\rangle_a} = i b_i^a b_j^a$. The $c$ Majorana fermions represent the matter sector, 
while the $b^a$ operators form a static $Z_2$ gauge field. In the flux-free sector, all bond operators take the value $+1$, resulting 
in a free fermion tight-binding model for the $c$ fermions.

For convenience, the $b$ operators may be combined into complex fermions defined on links, $\chi_{\langle ij\rangle_a} = (b_i^a + i b_j^a)/2$, 
in which case the bond operator becomes $u_{\langle ij\rangle_a} = 2 \chi_{\langle ij\rangle_a}^\dagger \chi_{\langle ij\rangle_a} - 1$. 
The flux-free sector corresponds to the vacuum of these fermions, where 
$\langle \chi_{\langle ij\rangle_a}^\dagger \chi_{\langle ij\rangle_a} \rangle = 1$ on all links.

\begin{figure}[h]
\centering \epsfig{file=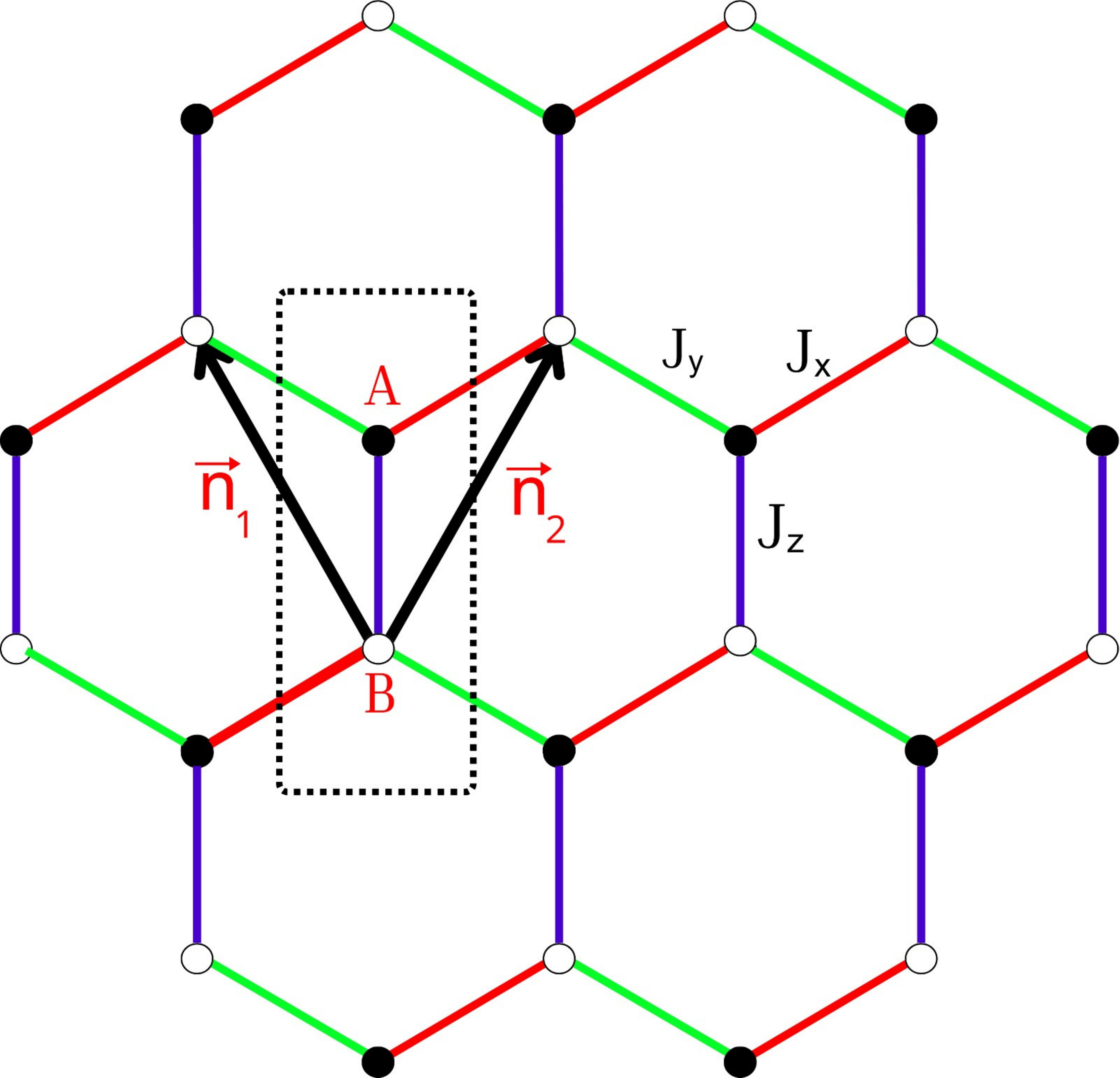,width=0.6\linewidth}
\caption{Honeycomb lattice hosting the exactly solvable Kitaev model, where spin interactions depend on bond direction.}
\label{fig.honeycomb}
\end{figure}

Analogously to the gauge sector, the matter sector is solved by defining complex fermions composed of $c$ Majoranas from different 
sublattices. As $c_i = c_i^\dagger$, complex fermions must be constructed from pairs of sites. We define $f_r = (c_{rA} + i c_{rB})/2$, 
where $r$ denotes the unit cell. While we choose the $z$-bond direction for this construction (see Fig.~\ref{fig.honeycomb}), the 
choice is arbitrary in the isotropic case, and physical results remain invariant under bond direction.

The Fourier transform of $H_K$ in terms of $f$ fermions yields
\begin{equation}
H_K = \sum_q (f_q^\dagger f_{-q}) \begin{pmatrix}
\zeta_q & i\Delta_q \\
-i\Delta_q & -\zeta_q
\end{pmatrix} \begin{pmatrix}
f_q \\
f_{-q}^\dagger
\end{pmatrix},
\end{equation}
where $\zeta_q = \text{Re} S_q$, $\Delta_q = \textrm{Im}S_q$, and $S_q = \sum_{\eta_a} J_a e^{i\bm{q} \cdot \bm{\eta}_a}$. 
The structure factor $S_q$ depends on the bond vectors $\bm{\eta}_x = (\sqrt{3}a/2, a/2)$, $\bm{\eta}_y = (-\sqrt{3}a/2, a/2)$, 
and $\bm{\eta}_z = (0, -a)$, where $a$ is the lattice spacing.

It is worth noting that the Kitaev model exhibits formal similarities with the BCS Hamiltonian, allowing 
it to be diagonalized through a Bogoliubov transformation of the form $f_q = \cos\psi_q, a_q + i\sin\psi_q, a_{-q}^\dagger$, 
with $\tan 2\psi_q = -\Delta_q/\zeta_q$. This yields $H_K = \sum_q E_q(a_q^\dagger a_q - 1/2)$, where the excitation spectrum is 
$E_q = 2|S_q| = \hbar \nu_q$ for the matter fermions. In the continuum limit, the momentum summation becomes an integral over the 
first Brillouin zone (BZ), whose area is $A_{BZ} = 8\sqrt{3}\pi^2 / 9a^2 \approx 15.2/a^2$, while $a\sim 10^{-10} m$. 
Since $c_q^\dagger = c_{-q}$, the $c$ fermions are not independent, and the sum is restricted to the half Brillouin zone (HBZ).

In the isotropic limit, the spectrum vanishes at six Dirac points located at the corners of the BZ, given by $q_c = \pm(4\pi/3\sqrt{3}a, 0)$ and $q_c = (\pm2\pi/3\sqrt{3}a, \pm2\pi/3a)$. Near these points, the structure factor behaves linearly:
\begin{equation}
S_q \approx \frac{3}{2} J a (q_x + i q_y),
\end{equation}
which corresponds to a relativistic dispersion $E_q = \hbar q c$ with effective speed $c = 3J a / \hbar$.

In the Kitaev model, the QSL exhibits a finite spin gap, and spin correlations are extremely short-ranged due to the presence of 
local $Z_2$ symmetries, in contrast to the power-law decay observed in the Heisenberg chain.

\section{Spin current evaluation}
\label{sec.spincurrent}

In the present configuration, the spin current flowing through the Kitaev/SC interface is intrinsically a nonequilibrium effect, generated by the 
finite chemical potential imbalance between spin-up and spin-down quasiparticles in the superconducting reservoir. Whenever $\Delta\mu\neq0$, 
the system is driven into a nonequilibrium steady state that sustains a finite, time-independent spin flux across the interface. In contrast, 
thermodynamic equilibrium would require $\mu_\uparrow=\mu_\downarrow$, implying the absence of spin currents. In this biased regime, spin is 
injected from the superconductor into the Kitaev layer through quasiparticle processes governed by the splitting $\Delta\mu$, while the 
anisotropic correlations of the Kitaev model determine how this injected spin couples to the emergent Majorana excitations. Formally, 
the transport can be treated within a linear-response framework, where $\Delta\mu$ acts as the generalized force driving the current. 
However, the finite spin bias explicitly breaks detailed balance, so the stationary current reflects the physics of a nonequilibrium 
steady state rather than equilibrium behavior. Assuming a small imbalance in the chemical potential, we employ linear response theory to 
evaluate the injected spin current, as detailed in this section.

As a result of spin-flip scattering, there is a reduction in the number of electrons with spin-up and an increase in those with 
spin-down at the interface. Subsequently, the spin current operator is defined as 
$I_s=(\hbar/2)\sum_k(\dot{n}_{k\downarrow}-\dot{n}_{k\uparrow})$, where $n_{k\sigma}=c_{k\sigma}^\dagger c_{k\sigma}$ represents
the electron number operator. To maintain the injected spin current, we presuppose a constant chemical potential imbalance, 
$\Delta\mu=\mu_\uparrow-\mu_\downarrow$, between the spin-up and spin-down electrons. Such an imbalance can be sustained, for instance, 
through spin accumulation engendered by spin-orbit coupling. Provided that the electron number operator commutes with both the superconducting 
and Kitaev Hamiltonians, given by Eq. (\ref{eq.Hsc}) and (\ref{eq.HK}), the time derivative of $n_{k\sigma}$ is determined solely by 
the commutator $[n_{k\sigma},H_{sd}]$. Consequently, the Heisenberg equation of motion provides a straightforward expression for the 
injected spin current, as expressed by 
\begin{equation}
    I_s=\frac{2i J_{sd}}{N_e}\sum_{ikp}\left[\sigma_i^+ c_{p\downarrow}^\dagger c_{k\uparrow}e^{i(\bm{k}-\bm{p})\cdot\bm{r_{i}}}\right]+h.c.
\end{equation}
In the interaction picture, the expected value of the spin current is derived from 
$\langle I_s(t)\rangle=\langle S^\dagger(t)\hat{I}_s(t) S(t)\rangle_0$, where the caret denotes time evolution in accordance with the 
free Hamiltonian $H_0=H_{SC}-H_K$, and $S(t)=T_t\exp[-(i/\hbar)\int_{-\infty}^t\hat{H}_{sd}(t^\prime)dt^\prime]$ represents the $S$ matrix. 
Additionally, the subscribed index in $\langle\ldots\rangle_0$ signifies that the average is computed utilizing the quadratic Hamiltonian $H_0$. 
For scenarios involving small coupling at the interface, we adopt the linear series expansion of the $S$ matrix, leading to 
\begin{equation}
\langle I_s\rangle=-\frac{i}{\hbar}\int_0^\infty \langle[\hat{I}_s(t),\hat{H}_{sd}(0)]\rangle_0dt.
\end{equation}

In the normal state, the expectation values involving two electronic annihilation or creation operators are null; however, 
in the superconducting phase, contributions to the injected spin current also arise from terms of the form $\langle c_{k\sigma}c_{-k-\sigma}\rangle_0$.
To derive an appropriate expression for the spin current, it is necessary to replace the electronic operators by the superconducting 
quasiparticle operators. Following a meticulous procedure, the expression for the spin current is given by 
\begin{align}
    \langle I_s\rangle=\frac{4J_{sd}^2}{N_e^2\hbar}\sum_{ijkp}\int_0^\infty\left[ \cos^2\Delta\theta_{kp}M_{ijkp}[\hat{A}](t)+\right.\nonumber\\
    \left.+\sin^2\Delta\theta_{kp}\frac{M_{ijkp}[\hat{B}](t)+M_{ijkp}[\hat{C}](t)}{2}\right]dt,
\end{align} 
where we define commutator mean value $M_{ijkp}[\hat{A}](t)=\langle[\hat{A}_{ikp}(t),\hat{A}_{jkp}^\dagger(0)]\rangle_0-\langle[\hat{A}_{ikp}^\dagger(t),\hat{A}_{jkp}(0)]\rangle_0$, with analogous terms inferred for $M_{ijkp}[\hat{B}](t)$ and $M_{ijkp}[\hat{C}](t)$. 
The mixed operators are represented as 
\begin{align}
    \hat{A}_{ikp}(t)&=b_{k\uparrow}^\dagger b_{p\downarrow}\hat{S}_i^-(t)e^{-i(\bm{k}-\bm{p})\cdot\bm{r}_i}e^{i(\Omega_\mu+\nu_{k\uparrow}-\nu_{p\downarrow})t}\\
    \hat{B}_{ikp}(t)&=b_{k\uparrow}^\dagger b_{-p\uparrow}^\dagger \hat{S}_i^-(t)e^{-i(\bm{k}-\bm{p})\cdot\bm{r}_i}e^{i(\Omega_\mu+\nu_{k\uparrow}+\nu_{p\uparrow})t}\\
    \hat{C}_{ikp}(t)&=b_{-k\downarrow} b_{p\downarrow}\hat{S}_i^-(t)e^{-i(\bm{k}-\bm{p})\cdot\bm{r}_i}e^{i(\Omega_\mu-\nu_{k\downarrow}-\nu_{p\downarrow})t},
\end{align} 
where $E_{k\sigma}=\hbar\nu_{k\sigma}$ denotes the energy of the superconducting quasiparticle, which is generated (annihilated) by the operator 
$b_{k\sigma}^\dagger$ ($b_{k\sigma}$), and $\Delta\mu=\hbar\Omega_u$. In the above equation, $\Delta\theta_{kp}=\theta_k-\theta_p$, while 
$\theta_k$ and $\theta_p$ represent the angles associated with the Bogoliubov transformation, and
\begin{align}
    \cos^2\Delta\theta_{kp}=\frac{1}{2}\left[1+\frac{\xi_k\xi_p+|\Delta|^2}{E_k E_p} \right].
\end{align}
In the scenario where the superconducting gap becomes null, both angles reduce to zero, thus reverting 
the spin current expression to the NM regime \cite{jmmm624.173024}.

It can be readily demonstrated that $M_{ijkp}[\hat{A}](t)=2\textrm{Re}\langle[\hat{A}_{ikp}(t),\hat{A}_{jkp}^\dagger(0)]\rangle_0$, 
which facilitates the expression of the injected spin current as 
\begin{equation}
    \label{eq.Is}
    \langle I_s\rangle=-8J_{sd}^2\textrm{Im}\tilde{\chi}(0),
\end{equation} 
wherein $\tilde{\chi}(\omega)=\int\chi(t)e^{i\omega t}dt$ represents the temporal Fourier transform of the susceptibility 
defined by $\chi(t)=N_e^{-2}\sum_{ijkp}\chi_{ijkp}(t)$, in which 
\begin{align}
    \chi_{ijkp}(t)=\frac{1}{i\hbar}\theta(t)\left[\cos^2\Delta\theta_{kp}\langle[\hat{A}_{ikp}(t),\hat{A}_{jkp}^\dagger(0)]\rangle_0+\right.\nonumber\\
    \left.+\frac{\sin^2\Delta\theta_{kp}}{2}\langle[\hat{B}_{ikp}(t),\hat{B}_{jkp}^\dagger(0)]+[\hat{C}_{ikp}(t),\hat{C}_{jkp}^\dagger(0)]\rangle_0\right].
\end{align}

In the low-temperature regime, where the superconducting gap is significantly large, the requisite chemical imbalance to sustain 
the spin current exceeds the permissible maximum. Consequently, as seen in the SC/FM junction model \cite{prb102.024412}, a null spin current 
is expected for $T\gtrsim 0$. Only within the regime of finite temperatures, $T\lesssim T_c$, where the superconducting gap is reduced, 
is the chemical potential adequate to supply the requisite energy for inducing electron spin-flip scattering processes at the interface.

To obtain an analytical expression for the injected spin current, it is convenient to express the susceptibility as
\begin{equation}
    \label{eq.chi}
    \chi_{ijkp}(t)=\frac{1}{i\hbar}\theta(t)[F_{ijkp}^{-+}(t)-F_{jikp}^{+-}(t)],    
\end{equation}
where 
$F_{ijkp}^{-+}(t)=\Lambda_{ijkp}^{-+}(t)+\Xi_{ijkp}^{-+}(t)+\Phi_{ijkp}^{-+}(t)$, with the correlation functions
\begin{align}
    \label{eq.correlation}
    \Lambda_{ijkp}^{-+}(t)&=\cos^2\Delta\theta_{kp}\langle \hat{A}_{ikp}(t)\hat{A}_{jkp}^\dagger(0)\rangle_0\\
    \Xi_{ijkp}^{-+}(t)&=\frac{1}{2}\sin^2\Delta\theta_{kp}\langle \hat{B}_{ikp}(t)\hat{B}_{jkp}^\dagger(0)\rangle_0\\
    \Phi_{ijkp}^{-+}(t)&=\frac{1}{2}\sin^2\Delta\theta_{kp}\langle \hat{C}_{ikp}(t)\hat{C}_{jkp}^\dagger(0)\rangle_0,
\end{align}
and a similar definition to $F_{jikp}^{+-}(t)$. Consequently, the Fourier transform results in (see Appendix (\ref{appendixA}) 
for more details)
\begin{equation}
    \label{eq.imchitilde}
    \textrm{Im}\tilde{\chi}_{ij}(0)=-\frac{\delta_{ij}}{2\hbar}(1-e^{-\beta\Delta\mu})\tilde{F}_{ii}^{-+}(0).
\end{equation}
Note that, as expected, the spin current vanishes in the absence of a chemical potential. In addition, the evaluation of 
$\textrm{Im}\tilde{\chi}_{ij}(0)$ involves the temporal Fourier transform of the correlation functions given by Eq. (\ref{eq.correlation}). 
The three correlation functions provide similar results, and it is sufficient to analyze the first one, expressed by
    \label{eq.Lambdaijkp}
    \begin{align}
        &\tilde{\Lambda}_{ij}^{-+}(0)=\left.\frac{1}{N_e^2}\sum_{kp}\int \Lambda_{ijkp}^{-+}(t) e^{i\omega t}dt\right|_{\omega\to 0}\nonumber\\
        &=\frac{1}{N_e^2}\sum_{kp}\cos^2\Delta\theta_{kp}e^{i(\bm{k}-\bm{p})\cdot\Delta\bm{r}}f(E_{k\uparrow})[1-f(E_{p\downarrow})]\times\nonumber\\
        &\times\int\delta(\nu_{k\uparrow}-\nu_{p\downarrow}+\Omega_\mu-\nu)\tilde{D}_{ij}^{-+}(\nu)d\nu,
    \end{align}    
where $f(E)=(e^{\beta E}+1)^{-1}$ is the Fermi-Dirac distribution, and $\tilde{D}_{ij}^{-+}(\nu)$
is the temporal Fourier transform of the spin correlation
\begin{equation}
    D_{ij}^{-+}(t)=\langle \hat{\sigma}_i^-(t)\hat{\sigma}_j^+(0)\rangle_0.
\end{equation}
It should be noted that the Dirac delta function ensures the conservation of energy during electron spin-flip scattering, and the same
outcome can be derived using the Golden Fermi rule. Additionally, $\hbar\nu$ represents the Majorana excitation energy on the Kitaev model.

To evaluate $\tilde{\Lambda}_{ij}(0)$, we convert the momentum into integrals, which are determined by using spherical
coordinates. The angular part of the $k$ integral provides
\begin{equation}
    \int_0^{2\pi}d\varphi\int_0^\pi  d\theta \sin\theta e^{ik\Delta r \cos\theta}=4\pi\textrm{sinc}(k\Delta r),
\end{equation}
where $\textrm{sinc}(x)=x^{-1}\sin x$, with a similar result from the $p$ integral. The superconducting state demands that 
$k\approx k_F\sim 10^{10}$ m$^{-1}$, and the $\textrm{sinc}(k_F\Delta r)$ becomes relevant only for $\Delta r\approx 0$. Indeed, adopting 
$\Delta r$ at an order consistent with the scale of the lattice spacing, we derive $\textrm{sinc}(k_F a)\sim 10^{-2}$, which justifies 
considering $\textrm{sinc}^2(k_F\Delta r)\approx\delta_{ij}$. Expressing the remainder radial integral in terms of the electronic energy, we obtain
\begin{align}
    \tilde{\Lambda}_{ij}^{-+}(0)=\frac{\hbar\delta_{ij}\rho_F^2}{4N_e^2}\int d\xi\int d\xi^\prime \frac{EE^\prime+|\Delta|^2}{2EE^\prime}\times\nonumber\\
    \times\int f(E_\uparrow)[1-f(E_\downarrow)]\delta(E-E^\prime-\hbar\nu)\tilde{D}_{ij}^{-+}(\nu)d\nu,
\end{align}
where $E=\sqrt{\xi^2+|\Delta|^2}$, and $E_\sigma=E-\sigma\Delta\mu/2$. Since the energy integrals are limited to the thin interval $[-\hbar\omega_D,\hbar\omega_D]$,
we have adopted a constant density of energy at the Fermi level $\rho_F=V_e mk_F/\pi^2\hbar^2=1.5N_e/\epsilon_F$. Employing the quasiparticle energy, the correlation
is given by
\begin{align}
   & \tilde{\Lambda}_{ij}^{-+}(0)=\frac{\hbar\delta_{ij}}{8N_e^2}\int d\nu\int d E \rho(E)\rho(E-\hbar\nu)\times\nonumber\\
   &\times\left[1+\frac{|\Delta|^2}{E(E-\hbar\nu)}\right]\tilde{D}_{ij}^{-+}(\nu)f(E_\uparrow)f(\hbar\nu-E_\downarrow),
\end{align}
where
\begin{equation}
    \rho(E)=\rho_F\sqrt{\frac{E^2}{E^2-|\Delta|^2}}
\end{equation}
establishes the density of states for the superconducting (SC) phase. It is important to note that the electronic energy is restricted to a narrow 
shell around the Fermi level, such that $|\epsilon_k - \mu_\sigma| \leq \hbar \omega_D$. This constraint implies that the corresponding quasiparticle 
energies satisfy $E \leq \sqrt{(\hbar \omega_D)^2 + |\Delta|^2}$. As a consequence, the relevant frequency range is limited to 
$\nu \lesssim \omega_D$, and higher-energy scattering processes with $\hbar \nu \gtrsim \omega_D$ lie outside the regime of applicability of BCS theory. 
This argument justifies restricting the integration domain for the frequency to $0 \leq \nu \lesssim \omega_D$. Additionally, it should be noted
that the SC gap is a parameter that depends on the temperature $T$ and the chemical potential imbalance $\Delta\mu$, and its value is determined 
self-consistently from Eq. (\ref{eq.gapsc}).

Summing the three correlation functions, we obtain
\begin{align}
    \label{eq.Fmp}
    &\tilde{F}^{-+}(0)=\frac{\hbar\rho_F^2}{16N_e^2}\int d\nu\frac{\hbar\nu-\Delta\mu}{e^{\beta(\hbar\nu-\Delta\mu)}-1}\tilde{D}^{-+}(\nu)W(\nu),
\end{align}
where $\tilde{D}^{-+}(\nu)=\sum_i\tilde{D}_{ii}^{-+}(\nu)$, and
\begin{align}
    W(\nu)=&\int dE\frac{\rho(E)\rho(E-\hbar\nu)}{\rho_F^2}\left[1+\frac{|\Delta|^2}{E(E-\hbar\nu)}\right]\times\nonumber\\
    &\times\frac{f(E+\Delta\mu/2-\hbar\nu)-f(E-\Delta\mu/2)}{\hbar\nu-\Delta\mu}
\end{align}
establishes a weight function that incorporates the electronic contribution to the spin current injection. 
At $T = T_c$, the SC gap vanishes and the function $W(\nu)$ reduces to a simpler form, denoted as $W_{NM}(\nu)$, which is approximately constant for
$\hbar \nu \gg \Delta_F$. In order to avoid unphysical divergences in the evaluation of $W(\nu)$, we substitute $E$ with $E+i\Gamma$, in which $\Gamma$ is the Dynes 
phenomenological parameter introduced to account for the effects of pair-breaking processes \cite{prb94.144508,prb97.014517}, and we take only the real part of $\rho(E + i\Gamma)$. 
Figure (\ref{fig.W}) illustrates the behavior of $W(\nu)$ for different temperatures and $\Delta\mu=2\Delta_F$. The lower bound on the energy is determined by the Kitaev energy gap, which 
demands the condition $\hbar\nu \geq \Delta_F$, as discussed in detail in the following section. 
At the critical temperature, the weight function characteristic of normal metal can be approximated using an exponential saturation curve, 
leading to an outcome of $W_{NM}(\nu)\approx 1.9(1-0.5)e^{-0.2\hbar\nu/\Delta_F}$, and yielding a relative root-mean-square error of $0.90$ percent.
\begin{figure}[h]
\centering \epsfig{file=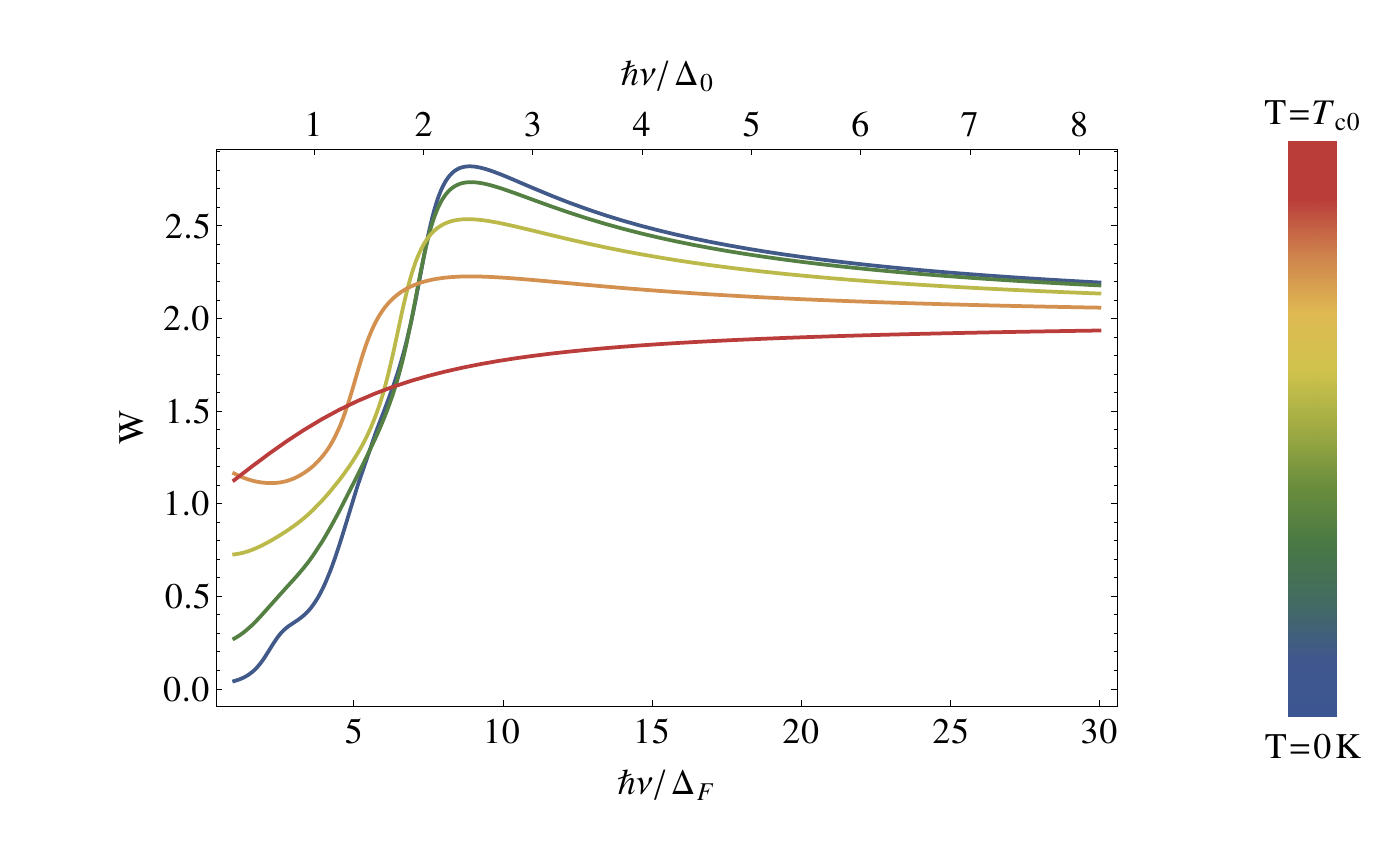,width=\linewidth}
\caption{The function $W(\nu)$ characterizes the electronic contribution to the correlation function $\tilde{F}^{-+}(0)$. In the high-energy limit, 
$\hbar \nu \gg \Delta_F$, this function asymptotically approaches the normal-metal value, $W_{NM}(\hbar \nu \gg \Delta_F) \approx 2$. Here, the critical temperature$(T_c$ denotes 
the SC transition temperature corresponding to $\Delta \mu = 2\Delta_F$; an analogous behavior is observed for other values of the chemical potential imbalance.}
\label{fig.W}
\end{figure}

Note that, in the normal phase, the energy $E_\sigma$ is replaced by $\xi_\sigma$, allowing the Fermi-Dirac distribution to be approximated by a step 
function near $\mu_\sigma$, due to the condition $\epsilon_\sigma\approx \epsilon_F\gg k_BT$ (at room temperature), which allows us to simplify 
the $\tilde{F}^{-+}(0)$ correlation function. On the other hand, in the SC state, $E_\sigma$ is of the same order as the superconducting gap, 
rendering the step function approximation inapplicable. 

Finally, it is noteworthy that the spin-spin correlation specified by $D_{ij}^{-+}(t)$ exhibits considerable generality, thereby enabling the application of 
the developed methodological framework to a diverse range of magnetic junctions beyond the Kitaev case. Replacing Eq. (\ref{eq.Fmp}) into Eq. (\ref{eq.Is}), we
obtain
\begin{align}
	\label{eq.Is2}
    \langle I_s\rangle&=\left(\frac{3J_{sd}}{4\epsilon_F}\right)^2(1-e^{-\beta\Delta\mu})\int d\nu\frac{\hbar\nu-\Delta\mu}{e^{\beta(\hbar\nu-\Delta\mu)}-1}\times\nonumber\\
    &\times \tilde{D}^{-+}(\nu)W(\nu).
\end{align}

To enable a direct comparison with the Kitaev model results, we compute the spin current in a conventional two‑dimensional FM defined on a hexagonal 
honeycomb lattice. It should be emphasized that, in the long-wavelength limit, the physical results are expected to be insensitive to the specific lattice geometry, 
apart from overall numerical prefactors. In order to stabilize long-range order along the $z$-axis at finite temperature, the Hamiltonian is endowed with an easy-axis 
anisotropy and assumes the conventional form $H_{\text{FM}} = -J \sum_{\langle ij \rangle} \left( S_i^x S_j^x + S_i^y S_j^y + \lambda S_i^z S_j^z \right)$,
where we use the same exchange coupling $J>0$ as before, $\lambda>1$ denotes the anisotropy parameter, and the sum is evaluated over the $z=3$ nearest neighbors. 
Despite the presence of this easy-axis term, the continuous $O(2)$ symmetry in the $xy$-plane remains preserved; consequently, in accordance with the Mermin–Wagner 
theorem, only the $S^z$ component of the spin acquires a finite expectation value for $0 \leq T \leq T_C$, where $T_C$ denotes the Curie temperature. 
Within the Holstein–Primakoff representation and in the linear spin-wave approximation, the Hamiltonian can be diagonalized as
$H_{\text{FM}} = \sum_q \varepsilon_q  a_q^\dagger a_q$,
where the magnon dispersion is given by $\varepsilon_q = \hbar \omega_q = 6 J S (\lambda - \text{Re}\,\gamma_q)$.  
The lattice structure of the honeycomb geometry enters via the complex structure factor 
$\gamma_q = (1/3)\left[ 2\cos\left(\sqrt{3} a q_x/2\right) e^{i a q_y/2} + e^{-i a q_y} \right]$.

In the long-wavelength regime $aq \ll 1$, the magnon dispersion reduces to a gapped, non-relativistic form,
$\varepsilon_q \approx \Delta_{\text{FM}} +1.5 J S a^2 q^2$, where $\Delta_{\text{FM}} = 6 J S (\lambda - 1)$ is the ferromagnetic gap 
induced by the easy-axis anisotropy. The on-site transverse spin–spin correlation function assumes the form
$D_{\text{FM}}^{-+}(t) = \sum_i D_{ii}^{-+}(t) = 4 S \sum_q  n_qe^{i \omega_q t}$,
where $n(\varepsilon_q) \equiv n_q = (e^{\beta \varepsilon_q} - 1)^{-1}$ denotes the Bose–Einstein distribution for magnons.  
Upon Fourier transforming to frequency space, one obtains the simpler result $\tilde{D}_{\text{FM}}^{-+}(\omega) = 8\pi S \sum_q n_q\, \delta(\omega + \omega_q)$,
which is the as the central input for evaluating the spin-current response for providing the spin current
\begin{align}
	\label{eq.IsFM}
    \langle I_s\rangle&\approx \left(\frac{J_{sd}}{8\epsilon_F}\right)^2 \frac{3N_u}{J} (1-e^{-\beta\Delta\mu})\int_{\Delta_{FM}}^\infty d\varepsilon W(-\varepsilon/\hbar)(\varepsilon+\nonumber\\
	&+\Delta\mu)e^{-\beta\varepsilon},
\end{align}
where $N_u$ denotes the total number of unit cells in the honeycomb lattice, and we restrict our analysis to the low-temperature regime characterized by $T \ll T_C$.
In the following analysis, we adopt $\lambda=1.1$, $a=\SI{5}{\angstrom}$, and $S=1/2$, which results in $\Delta_{FM}\approx1.15\Delta_F$.

The next section provides a detailed evaluation of the spin-spin correlation for the Kitaev model.

\section{Kitaev spin-spin correlation}
\label{sec.kitaev_ssc}
The evaluation of spin current requires the utilization of spin-spin correlation 
$D_{ij}^{-+}(t)=\langle \sigma_i^x(t)\sigma_j^x(0)\rangle_0+\langle \sigma_i^y(t) \sigma_j^y(0)\rangle_0$. As elucidated by 
Baskaran \cite{prl98.247201}, the Kitaev model manifests finite spin correlation only between spins located on the same 
link. For the unit cell defined along the $z$-link, the pertinent spin-spin correlations are specified by 
$\langle \sigma_{rA}^z(t)\sigma_{rA}^z(0)\rangle_0$, $\langle \sigma_{rA}^z(t)\sigma_{rB}^z(0)\rangle_0$, $\langle \sigma_{rB}^z(t)\sigma_{rA}^z(0)\rangle_0$, 
and $\langle \sigma_{rB}^z(t)\sigma_{rB}^z(0)\rangle_0$, where $r$ designates the unit cell position. Furthermore, 
due to the locality obtained from $\textrm{sinc}(k_F\Delta r)$, our investigation will be restricted to local correlations.
In the isotropic limit, this correlation is independent of any specific link orientation, thus permitting it to be represented as 
$D_{ii}^{-+}(t)=2D_r^{zz}(t)=2\langle 0| \sigma_r^z(t)\sigma_r^z(0)|0\rangle$. Here, the state $|0\rangle=|M_0\rangle|F_0\rangle$ denotes the 
vacuum state for the matter ($|M_0\rangle$) and flux sector ($|F_0\rangle$).

In terms of the Majorana fermions, the spin-spin correlation for sublattice A is expressed as 
$D_{rA}^{zz}(t)=\langle 0| ic_{rA}(t)b_{rA}^z(t)ic_{rA}(0)b_{rA}^z(0)|0\rangle$. Assuming the ground state is defined by the 
free-flux sector, we arrive at the results presented in $\chi_{rz}^\dagger \chi_{rz}|F_0\rangle=|F_0\rangle$ and $\chi_{rz}^\dagger|F_0\rangle=0$. 
Consequently, by utilizing complex fermions as described by $f_r$ and $\chi_{rz}$, the correlation is given by 
\begin{align}
    &D_{rA}^{zz}(t)=-\langle M_0|\langle F_0|e^{iH_Kt/\hbar}(f_{r}+f_r^\dagger)\chi_{rz}^\dagger e^{-iH_Kt/\hbar}(f_r+\nonumber\\
    &+f_r^\dagger)\chi_{rz}|F_0\rangle|M_0\rangle.
\end{align}

In accordance with Knolle \cite{knolle}, we employ the commutation relation $\chi_{rz}^\dagger e^{-iH_K t/\hbar}=e^{-i(H_K+V_z)t/\hbar}\chi_{rz}^\dagger$, 
wherein $V_z=-2iJ c_ic_j$ denotes a potential term involving spins along the $\langle ij\rangle_z$ link$  $ to 
decouple the flux and matter operators \cite{prl112.207203}. This potential term alters the flux on the two adjacent plaquettes that share the $\langle ij\rangle$ 
link, warranting the definition of $H_z=H_K+V_z$ as the pair flux Hamiltonian. Consequently, we derive 
\begin{align}
    &D_{rA}^{zz}(t)=-e^{iE_0 t/\hbar}\langle M_0| e^{-iH_z t/\hbar}[f_{r}(t)+f_r^\dagger(t)](f_r+\nonumber\\
    &+f_r^\dagger)|M_0\rangle\langle F_0|\chi_{rz}^\dagger\chi_{rz}|F_0\rangle,
\end{align} 
where $f_r(t)=e^{iH_zt/\hbar}f_r e^{-iH_zt/\hbar}$. The mean value defined above characterizes a quantum quench, akin to the x-ray edge 
problem \cite{prl98.247201}. In this scenario, a Majorana fermion is initially generated at $t=0$, followed by the abrupt introduction of a 
flux pair into the system. Subsequently, the Majorana fermions evolve according to the $H_z$ Hamiltonian until they are annihilated at time $t$. 
Within the adiabatic approximation, which is justified in the regime of low-energy excitations, it is considered appropriate to replace the 
ground state of $H_K$ with that of $H_z$ \cite{prl112.207203}. Therefore, the spin-spin correlation is expressed as
\begin{align}
    D_{rA}^{zz}(t)&\approx-e^{-i \Omega_F t}\langle [f_{r}(t)+f_r^\dagger(t)](f_r+f_r^\dagger)\rangle_z,
\end{align}
where the $z$ index corresponds to averages calculated utilizing the eigenstates of $H_z$, and $\Delta_F=\hbar\Omega_F\approx0.26J$ 
represents the necessary energy to create the flux pair \cite{ap321.2}. An analogous method applied to sublattice B yields
\begin{align}
    D_{rB}^{zz}(t)&\approx-e^{-i \Omega_F t}\langle [f_{r}(t)-f_r^\dagger(t)](f_r-f_r^\dagger)\rangle_z.
\end{align}
By summing over all lattice sites, as necessary for the determination of susceptibility, we obtain the expression
\begin{align}
    \tilde{D}^{zz}(\nu)&=2i\hbar\sum_r\int[G_r^>(t)-G_r^<(-t)]e^{i(\nu-\Omega_F)t}dt\nonumber\\
    &=2i\hbar\sum_r[\tilde{G}_r^>(\nu-\Omega_F)-\tilde{G}_r^<(-\nu+\Omega_F)],
\end{align}
where the greater and lesser-than Green's functions are provided by $\hbar G_r^>(t)=-i\langle f_r(t)f_r^\dagger(0)\rangle_z$ 
and $\hbar G_r^<(t)=i\langle f_r^\dagger(0)f_r(t)\rangle_z$, respectively.

One could proceed with the greater and lesser-than Green's functions evaluations; however, the determination of 
retarded and advanced Green's functions is generally more direct, thereby justifying the representation of 
the spin-spin correlation through references $\hbar G_{\textrm{ret},r}(t)=-i\theta(t)\langle\{f_r(t),f_r^\dagger(0)\}\rangle_z$,
and $\hbar G_{\textrm{adv},r}(t)=i\theta(-t)\langle\{f_r(t),f_r^\dagger(0)\}\rangle_z$. 
Given $\tilde{G}_{\textrm{ret},r}(\nu)-\tilde{G}_{\textrm{adv},r}(\nu)=\tilde{G}_r^>(\nu)-\tilde{G}_r^<(\nu)$, and 
$\tilde{G}_r^<(\nu)=-e^{-\beta\hbar\nu}\tilde{G}_r^>(\nu)$, we obtain 
$\tilde{G}_r^>(\nu)=[1-f(\hbar\nu)][\tilde{G}_{\textrm{ret},r}(\nu)-\tilde{G}_{\textrm{adv},r}(\nu)]$. 
In a similar manner, $\tilde{G}_r^<(-\nu)=[1-f(\hbar\nu)][\tilde{G}_{\textrm{adv},r}(-\nu)-\tilde{G}_{\textrm{ret},r}(-\nu)]$, 
and thus the required spin-spin correlation is simplified to 
\begin{align}
    \label{eq.Dmp}
    &\tilde{D}^{-+}(\nu)=-8\hbar f(\Delta_F-\hbar\nu)\sum_r\textrm{Im}[\tilde{G}_{\textrm{ret},r}(\nu-\Omega_F)+\nonumber\\
    &+\tilde{G}_{\textrm{ret},r}(-\nu+\Omega_F)].
\end{align}
Disregarding the condition specified in $T\lesssim T_c$ momentarily, one observes that at zero temperature, it is possible to replace
$f(\Delta_F-\hbar\nu)=1-f(\hbar\nu-\Delta_F)$ by the step function $\theta(\hbar\nu-\Delta_F)$. In this scenario, a 
minimum energy $\Delta_F$ is necessary to sustain 
the injected spin current. The situation bears resemblance to those encountered in superconducting tunneling, and  $\Delta_F$ (the energy for creating  
a two-flux excitation) acts as the superconducting gap. From this analysis, at finite temperatures, minor contributions to the spin current could be
reached for energies $\hbar\nu\lesssim\Delta_F$; notwithstanding, the bound enforced by $Z_{SC}$ is normally greater than $\Delta_F$. In Fig. 
(\ref{fig.Dmp}), the dependence of the surface Green's function density $\tilde{D}^{-+}/N_u$ on $\hbar\nu$ is illustrated, revealing 
the condition $\hbar\nu \gtrsim \Delta_F$.

\begin{figure}[h]
\centering \epsfig{file=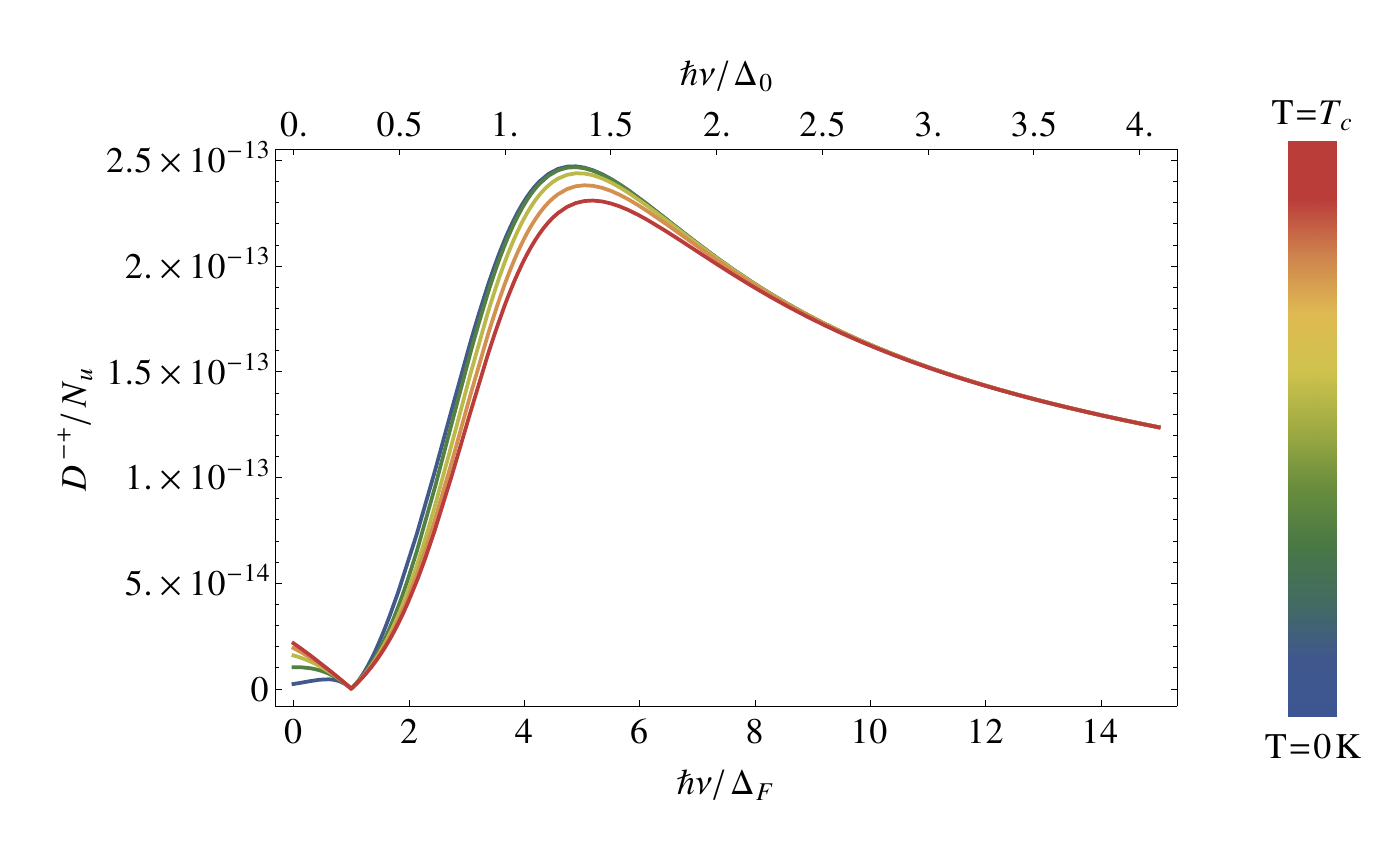,width=\linewidth}
\caption{The Majorana Green's function $\tilde{D}^{-+}(\nu)$ yields significant contributions exclusively for 
$\hbar\nu\geq\Delta_F$. $N_u$ denotes the total number of unit cells, and we employ the established value $\Delta_F = 0.26J$, and 
$T_c$ is defined as the SC critical temperature for $\Delta\mu=2\Delta_F$.}
\label{fig.Dmp}
\end{figure}

As demonstrated in Appendix (\ref{appendixB}), the retarded Green's function $\tilde{G}_{\textrm{ret},r}(\nu)$ 
exhibits spatial invariance, being given by
\begin{equation}
    \tilde{G}_\textrm{ret}(\nu)=\frac{\tilde{G}_\textrm{ret}^0(\nu)}{1-g_K\tilde{G}_\textrm{ret}^0(\nu)},
\end{equation}
In contrast, the unperturbed Green's function is described by
\begin{equation}
    \label{eq.G0retexact}
    \tilde{G}_\textrm{ret}^0(\nu)=\frac{1}{N_u}\sum_q \frac{h\nu+2\textrm{Re}(S_q)+i\varepsilon}{(\hbar\nu+i\varepsilon)^2-E_q^2},
\end{equation}
where $\varepsilon$ denotes an infinitesimal parameter originating from the analytical continuation of the Matsubara Green's function, 
and $g_K=-4J$ represents an interaction coupling constant. Note that, given that the unit cell area $A_{uc}=(2\pi)^2/A_{BZ}$, the number of unit cells 
within the interfacial area $A$ can be expressed as $N_u=A A_{BZ}/(2\pi)^2$.
\begin{figure}[h]
\centering \epsfig{file=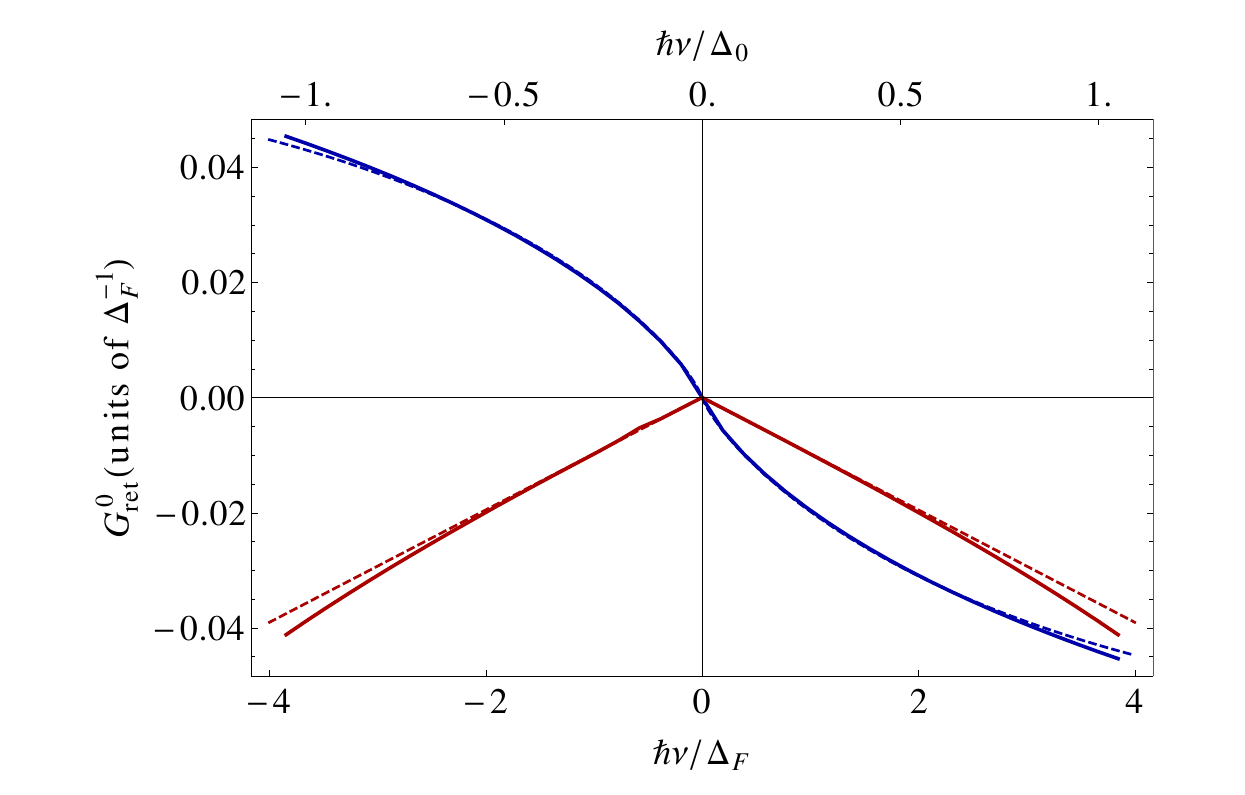,width=\linewidth}
\caption{The approximate retarded Green’s function (dashed line) is compared with the exact result (solid line), obtained via numerical integration. 
The red curves indicate the real part, while the blue curves represent the imaginary part of \(G_\textrm{ret}^0\). 
The cutoff parameter $Q = 2.1a^{-1}$ is chosen to yield an optimal fit between the approximate and exact solutions.}
\label{fig.G0ret}
\end{figure}

An exact evaluation of $\tilde{G}_\textrm{ret}^0(\nu)$ is unfeasible. However, given that the main contribution originates near the points 
that satisfy $E_q=0$, one can utilize the linear expansion around the vertices of the first Brillouin zone. It is noteworthy that each vertex 
is shared among three neighboring hexagons, resulting in each corner encompassing half of the first Brillouin zone. 
Therefore, the Green's function is simplified into 
\begin{equation}
    \tilde{G}_\textrm{ret}^0(\nu)=\frac{2\pi}{A_{HBZ}}\int_0^Q\frac{(\hbar\nu+i\varepsilon)q}{(\hbar\nu+i\varepsilon)^2-\hbar^2q^2c^2}dq,
\end{equation}
wherein $Q$ serves as a wave number cutoff approximately of the order of $\pi/a$, which is employed as a variational parameter. 
The value of $Q$ is chosen by minimizing the difference between the approximated and the exact Green's function. In typical cases, 
subsequently verified, $|\hbar\nu|\lesssim \Delta_F\ll \hbar Qc$. Over integration, the limiting condition $\varepsilon\to 0$ 
yields the straightforward result 
\begin{equation}
    \label{eq.G0ret}
    \tilde{G}_\textrm{ret}^0(\nu)\approx\frac{2\pi}{\hbar c^2 A_{BZ}}\left[2\nu\ln\left(\frac{|\nu|}{Qc}\right)-i\pi |\nu|\right].
\end{equation}
A thorough analysis indicates that $Q=2.1a^{-1}$ provides the most precise cutoff, derived from the numerical integration of 
Eq. (\ref{eq.G0retexact}). It is important to note that the real (imaginary) component 
of $\tilde{G}_\textrm{ret}^0(\nu)$ exhibits odd (even) symmetry in relation to the frequency $\nu$. Fig. (\ref{fig.G0ret}) 
demonstrates both the exact retarded Green's function and the approximate retarded Green's function.

The parity properties of the real and imaginary components of $\tilde{G}_\textrm{ret}^0(\nu)$ result in  
\begin{equation}
    \textrm{Im}[\tilde{G}_\textrm{ret}(\nu)+\tilde{G}_\textrm{ret}(-\nu)]=2Z(\nu)\textrm{Im}\tilde{G}_\textrm{ret}^0(\nu),
\end{equation}
where $Z(\nu)$ is a renormalization factor given by
\begin{equation}
    Z(\nu)=\frac{1+g_K^2|\tilde{G}_\textrm{ret}^0(\nu)|^2}{[1+g_K^2|\tilde{G}_\textrm{ret}^0(\nu)|^2]^2-4g_K^2[\textrm{Re}\tilde{G}_\textrm{ret}^0(\nu)]^2}.
\end{equation}
Note that, when $g_K$ vanishes, $Z(\nu)=1$ and the non-interacting results are recovered. Finally, we obtain the desired result
for the Kitaev spin-spin correlation,
\begin{align}
    \label{eq.D}
    &\tilde{D}^{-+}(\nu)=\frac{8A}{c^2}f(\Delta_F-\hbar\nu)Z(\nu-\Omega_F)|\nu-\Omega_F|\nonumber\\
    &=\frac{4\sqrt{3}}{3}\frac{\hbar}{J^2}f(\Delta_F-\hbar\nu)Z(\nu-\Omega_F)|\hbar\nu-\Delta_F| N_u,
\end{align}
which is independent of the lattice parameter $a$.

\section{Spin current and conductance}
\label{sec.results}
In a NM/FM junction, the interfacial spin current $J_s$ quantifies the flow of spin angular momentum exchanged between itinerant electrons in the 
NM and the magnetization of the FM. This exchange is governed by the spin-mixing conductance $g^{\uparrow\downarrow}$, which characterizes the 
efficiency with which transverse spin components are absorbed at the interface\cite{rezende}. When a spin accumulation $\Delta\boldsymbol{\mu}$ is present in the NM, the 
spin current injected into the FM (spin-transfer torque) is given by
\begin{align}
\mathbf{J}_s= \frac{g_r^{\uparrow\downarrow}}{4\pi}\mathbf{m}\times(\Delta\boldsymbol{\mu}\times\mathbf{m})+\frac{g_i^{\uparrow\downarrow}}{4\pi}(\Delta\boldsymbol{\mu}\times\mathbf{m}),
\end{align}
where $ \mathbf{m} $ is the unit vector along the magnetization direction, and $g_r^{\uparrow\downarrow}$ ($g_i^{\uparrow\downarrow}$) denotes the 
real (imaginary) part of the spin-mixing conductance. For a wide class of materials, $g_r^{\uparrow\downarrow}\gg g_i^{\uparrow\downarrow}$ only the first 
term provides a significant contribution, while the imaginary component can typically be neglected.
Conversely, a time-dependent magnetization $\dot{\mathbf{m}}\neq 0$ pumps a spin current proportional to $g_r^{\uparrow\downarrow}\mathbf{m}\times\dot{\mathbf{m}}$ 
into the NM, demonstrating that spin-transfer torque and spin pumping are reciprocal processes controlled by the same interfacial parameter $g^{\uparrow\downarrow}$. 
Typical values for spin-mixing and spin accumulation include $g_r^{\uparrow\downarrow} \sim 10^{18}\,\mathrm{m^{-2}}$ and $\Delta\mu \sim 10^{-4}\,\mathrm{eV}$
\cite{prl107.066604,jap111.07c307}. Therefore, a straightforward order-of-magnitude estimate yields a spin current density 
$J_s \propto \frac{1}{4\pi}\,g_r^{\uparrow\downarrow}\,\Delta\mu \sim 10^7 (\hbar/2e)\,\mathrm{A\,m^{-2}}$, when we 
include extra attenuation from non-ideal interfaces, spin backflow, and incomplete polarization.

The above relation between the injected spin current and the magnetization direction does not apply to the Kitaev scenario because its underlying 
assumptions are violated. This expression relies on a semiclassical description in which the ferromagnet is characterized by the well-defined, slowly varying order 
parameter $\mathbf{m}$, and the interfacial spin transfer is described as the scattering of itinerant electrons off a static exchange field. In contrast, the Kitaev model describes a 
system of strongly correlated, localized spins with highly anisotropic bond-dependent interactions and, in its quantum spin-liquid regime, no long-range magnetic order at all. 
Consequently, there is no macroscopic magnetization vector that can serve as a reference axis for defining transverse spin absorption or spin-transfer torque. Moreover, the spin 
degrees of freedom fractionalize into emergent quasiparticles (Majorana fermions and gauge fluxes), so angular momentum transfer at an interface cannot be captured by a single 
spin-mixing conductance $g^{\uparrow\downarrow}$.
\begin{figure}[h]
\centering \epsfig{file=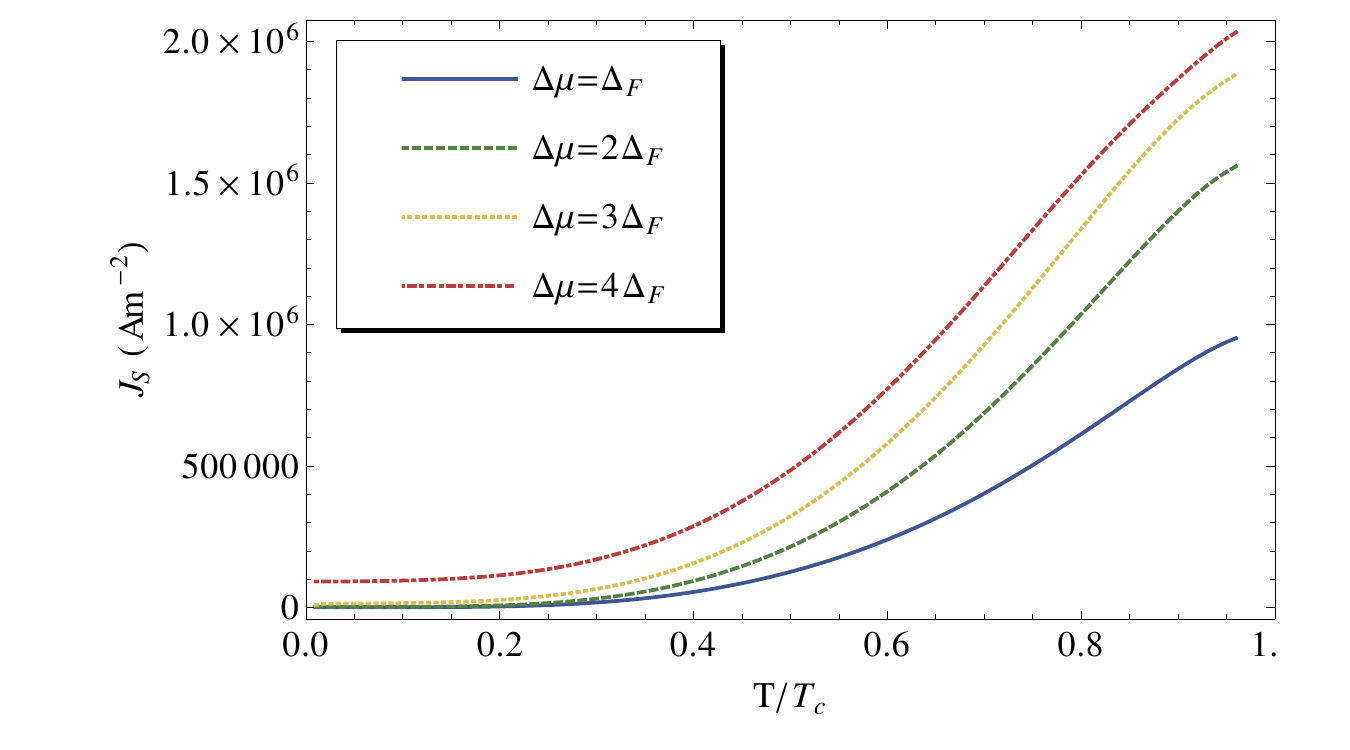,width=\linewidth}
\caption{The temperature dependence of the spin current injected into the Kitaev model from the SC layer is governed by the presence of the superconducting energy gap. 
As a consequence, the spin current is strongly suppressed and becomes effectively negligible in the low-temperature regime. A qualitatively similar temperature dependence is observed in 
a FM/SC junction; however, in this case the magnitude of the spin current is reduced by up to two orders compared with the Kitaev-based configuration. Here, the critical temperature $T_c$ 
is defined individually for each curve as a function of the corresponding value of $\Delta \mu$. The spin current is expressed in units of $\hbar/2e$.}
\label{fig.JsK}
\end{figure}

To properly determine the injected spin current in the Kitaev/SC junction, we substitute Eq. (\ref{eq.D}) 
into Eq. (\ref{eq.Is2}), which allows us to express the spin current density $J_s=\langle I_s\rangle/A$ as
\begin{align}
    \label{eq.JsK}
    J_s&=\frac{\hbar}{2}\left(\frac{J_{sd}}{a J \epsilon_F}\right)^2(1-e^{-\beta\Delta\mu})\int \frac{\hbar\nu-\Delta\mu}{e^{\beta(\hbar\nu-\Delta\mu)}-1}\times\nonumber\\
    &\times\frac{|\Delta_F-\hbar\nu|}{e^{\beta(\Delta_F-\hbar\nu)}+1}W(\nu)Z(\nu)d\nu,
\end{align}
where, in the following analysis, the lattice constant is set to $\SI{5}{\angstrom}$.

\begin{figure}[h]
\centering \epsfig{file=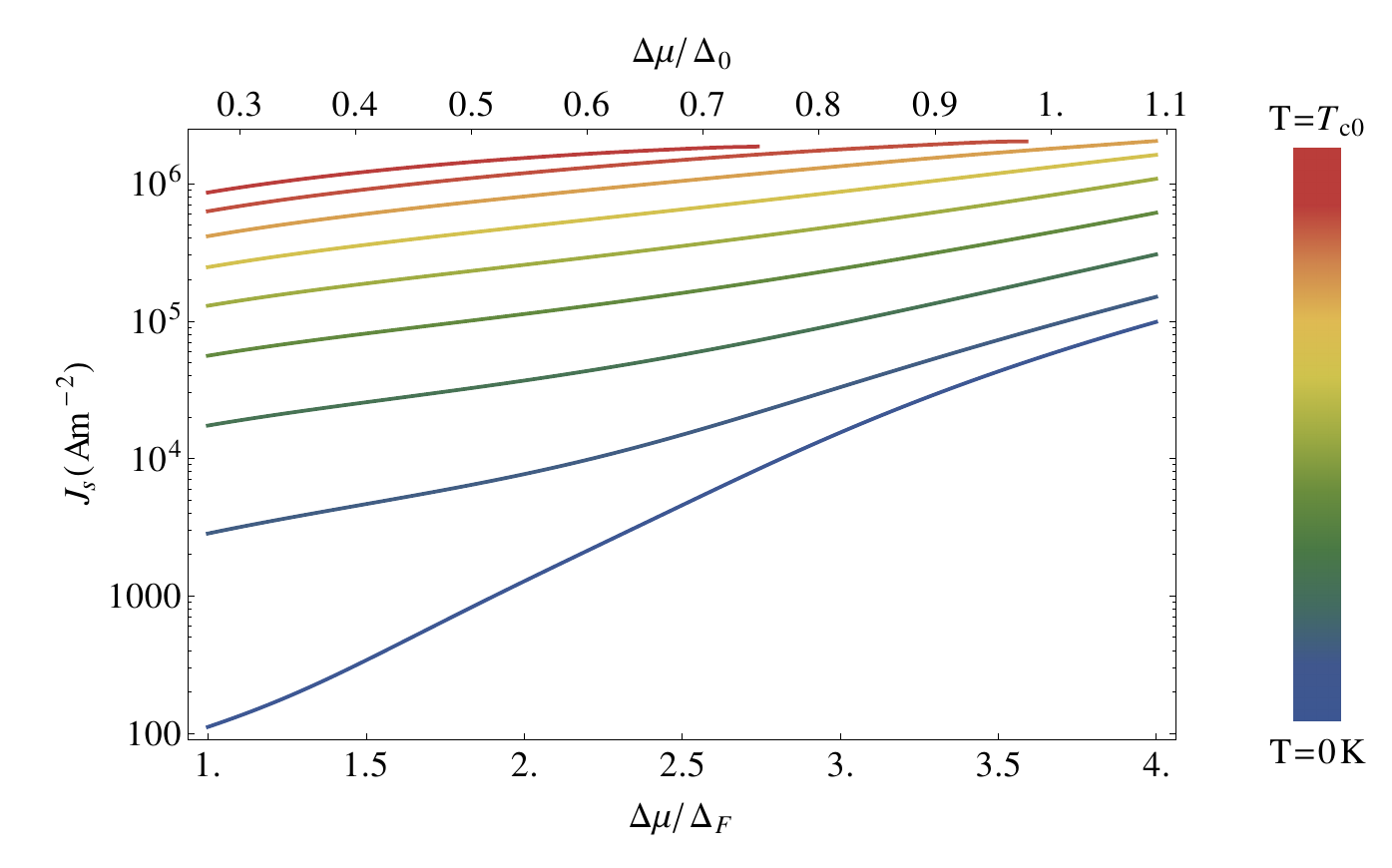,width=\linewidth}
\caption{The dependence of the spin current on the chemical imbalance exhibits an approximately exponential behavior, as evidenced by the linear trend 
in the logarithmic representation. The spin current is expressed in units of $\hbar/2e$.}
\label{fig.logJsK}
\end{figure}

Fig. (\ref{fig.JsK}) shows the evaluated spin current across the Kitaev/SC junction for some values of the chemical imbalance, as obtained from Eq. (\ref{eq.JsK}). 
The FM/SC configuration displays a qualitatively similar trend; however, in contrast to the Kitaev/SC interface, the corresponding spin current becomes negligible for 
temperatures $T \lesssim  0.2\,T_c$. The absolute magnitude of the spin current in these Kitaev/SC structures is approximately one order of magnitude smaller than the typical values 
reported for conventional FM/NM interfaces. Nevertheless, when directly compared with the FM/SC case, the Kitaev model can yield spin currents up to two orders of magnitude larger 
than those observed in FM/NM systems. As expected, in the low-temperature regime, the magnon population is strongly suppressed, which significantly reduces the spin current. 
In spite of this, the fact that spin transport in the Kitaev model is mediated by QSL excitations leads to an overall enhancement of spin conduction across the junction. The dependence 
on the chemical imbalance is shown in Fig. (\ref{fig.logJsK}). As can be observed, the curve is approximately linear, indicating that the spin current exhibits an exponential 
dependence on $\Delta\mu$.  Specifically, for $T = 0.6T_{c0}$, a linear regression yields the following expression for the spin current: 
$J_s(\Delta\mu) = \exp(11.865 + 0.606\,\Delta\mu/\Delta_F) (\hbar/2e)\si{\ampere\meter^{-2}}$, which corresponds to a relative root-mean-square error of only 0.6\% over the interval 
$\Delta_F \leq \Delta\mu \leq 4\Delta_F$.

Typically, the detection of a spin current is achieved by its conversion into a charge current via the inverse spin Hall effect in an adjacent metallic layer.
In this scenario, it is expected that when a spin current is injected from the Kitaev layer, only a fraction of the interfacial spin angular momentum flux contributes to the 
electrically detected signal \cite{nature410.345,prb66.060404,prl88.236601,apl88.182509}. Several loss mechanisms reduce the effective spin current. First, 
interfacial spin de-phasing and spin-memory loss, arising from interfacial disorder, orbital hybridization, and spin-orbit coupling, lead to partial 
randomization of the spin polarization already at the interface. Second, spin backflow occurs when the finite spin accumulation in the metal drives a counter-flow 
of spin current back into the ferromagnet, thereby reducing the net transmitted spin flux; this effect is commonly accounted for through a reduced effective 
spin-mixing conductance. Third, within the metallic layer, the injected spin current undergoes diffusive relaxation over the spin diffusion length, so that 
a significant portion of the spin angular momentum is dissipated before reaching the region where spin-to-charge conversion takes place. Finally, the conversion of spin 
current into a measurable charge current, for instance via the inverse spin Hall effect, is itself intrinsically inefficient and limited by the spin Hall angle of the metal. 
As a result of these combined processes, the spin current inferred from electrical measurements typically represents only a small fraction, often at the 
percent level, of the total spin current initially transmitted across the interface.
\begin{figure}[h]
\centering \epsfig{file=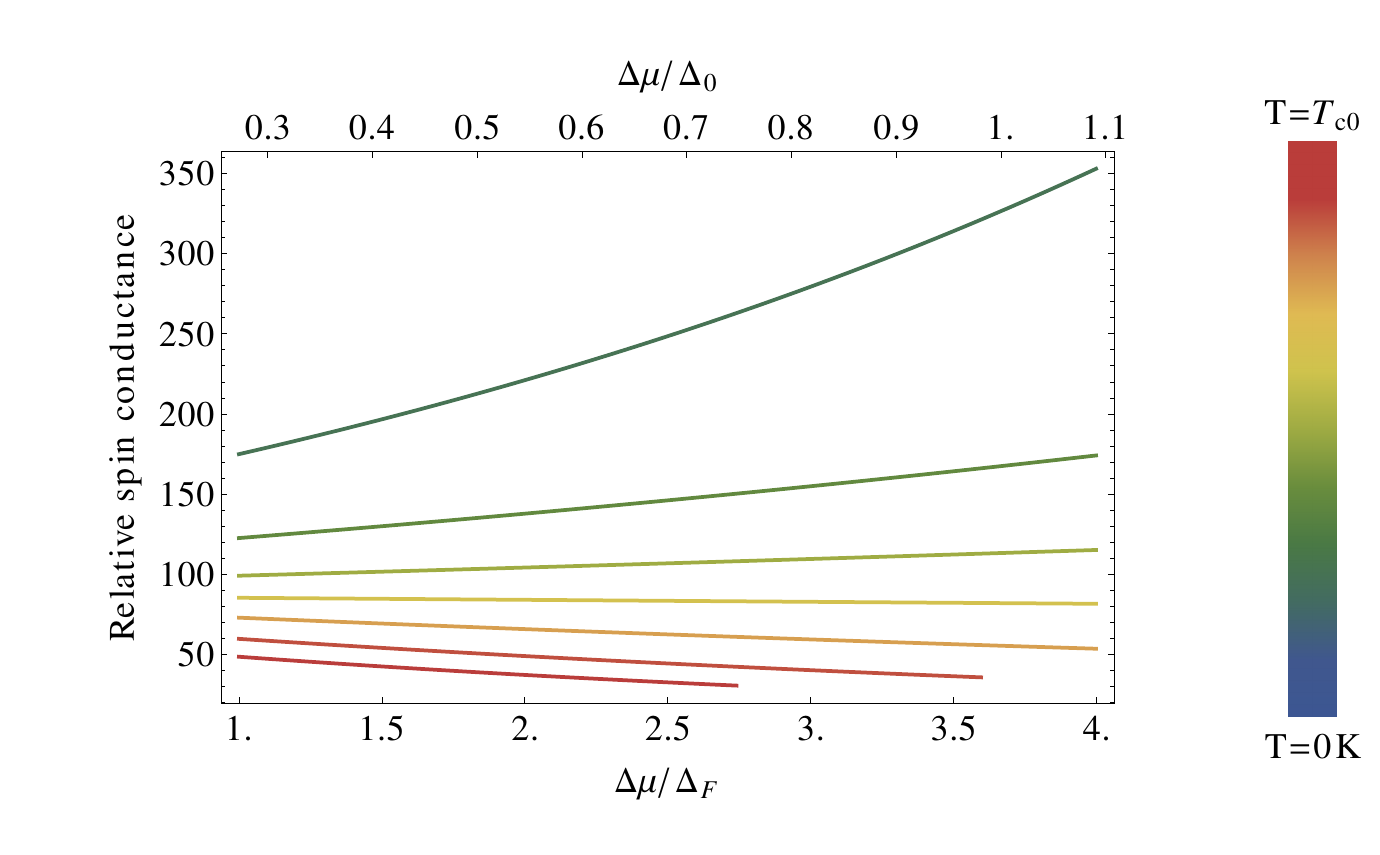,width=\linewidth}
\caption{The dependence of the relative spin conductance, $g_K/g_{FM}$, on the chemical imbalance is shown for several temperatures between absolute zero and $T_{c0}$. 
In the very-low-temperature regime, the Kitaev/SC interface exhibits significantly enhanced spin transport compared with the corresponding FM configuration.}
\label{fig.G}
\end{figure}

The spin conductance across the interface is defined as $g=\partial\langle J_s\rangle/\partial\Delta\mu$. It is important to consider that the chemical imbalance 
is confined to $\Delta_F\leq\Delta\mu \leq \Delta\mu_\textrm{max}$, with the lower boundary imposed by the presence of the Kitaev gap and the upper boundary established 
by the constraint $\Delta\mu\leq 1.20\Delta_0\approx 4.4\Delta_F$. Figure (\ref{fig.G}) displays the relative spin conductance as a function of temperature within the relevant 
range of chemical imbalances. The relative conductance is defined as the ratio $g_K/g_{FM}$. For temperatures $T \lesssim 0.2 T_{c0}$, the spin current through the FM junction 
becomes negligible, and consequently, the relative conductance $g$ exceeds by several orders of magnitude the values reported in Fig. (\ref{fig.logJsK}). As expected, for temperatures 
approaching absolute zero, only QSL excitations are capable of transporting spin current.

\section{Summary and conclusions}
\label{sec.conlusion}
In this study, we examine the injected spin current through a non-conventional interface consisting of a superconductor 
coupled with a Kitaev spin liquid. Although pure Kitaev materials have not yet been observed experimentally, there is substantial expectation 
that the ruthenium-based compound $\alpha$-\ce{RuCl_3} provides the necessary conditions to potentially confirm the existence of a two-dimensional 
QSL state. Under these circumstances, several authors have proposed using indirect measurements, particularly those involving spin transport, to 
confirm the presence of a QSL. Given the requirement for low-temperature experimental conditions, it is logical to consider a superconductor 
material interfacing with a Kitaev model, which demands the investigation of spin transport in Kitaev/superconductor junctions. 
Through a comprehensive theoretical analysis, we investigate the mechanism of spin current injection into a genuine QSL material. 
In this scenario, spin transport is facilitated by the Majorana fermions intrinsic to the Kitaev model. Our findings reveal that, contrary 
to the conventional scenario involving a NM/FM junction, the injected spin current is confined by the restricted chemical 
imbalance $\Delta_F\leq \Delta\mu\leq \Delta\mu_\textrm{max}$, with the boundaries delineated by the energy gaps of the Kitaev model (lower boundary) and 
the superconductor (upper boundary). Furthermore, the presence of the SC energy gap reduces the density of available quasiparticle states that can 
participate in spin-flip scattering processes, thereby leading to a suppression of the injected spin current, mainly in the case of conventional ferromagnetic layers. 
In the very–low-temperature regime, conventional materials are unable to support efficient spin transport because the population of magnonic carriers becomes 
strongly suppressed. In contrast, the behavior of the Kitaev/SC junction is markedly different. Owing to the presence of QSL excitations, the Kitaev/SC interface can 
sustain substantial spin transport even at temperatures approaching absolute zero. Finally, while the primary focus of this study is the SC/Kitaev interface, 
the formalism developed herein is sufficiently general. With minimal alterations, the derived results can be extended to encompass various interfaces, 
thereby broadening the scope of research into other QSL states.\\

This study was financed in part by the Coordenação de Aperfeiçoamento de Pessoal de Nível Superior – Brasil 
(CAPES) - Finance Code 001, and by the National Council for Scientific and Technological Development – CNPq. 

\appendix
\section{Susceptibility} 
\label{appendixA}
The induced spin current depends on the imaginary part of the temporal Fourier transform 
\begin{equation}
    \tilde{\chi}_{ijkp}(\omega)=\int \chi_{ijkp}(t)e^{i\omega t}dt,
\end{equation}
which can be expressed as 
\begin{align}
    &\textrm{Im}\tilde{\chi}_{ijkp}(\omega)=\frac{1}{2i}\int[\chi_{ijkp}(t)-\bar{\chi}_{ijkp}(-t)]e^{i\omega t}dt\nonumber\\
    &=\frac{1}{2\hbar}\int\{\theta(t)[F_{jikp}^{+-}(t)-F_{ijkp}^{-+}(t)]+\theta(-t)[F_{ijkp}^{+-}(t)-\nonumber\\
    &-F_{jikp}^{-+}(t)] \}e^{i\omega t}dt,
\end{align} 
where we have used the relation $\bar{F}_{ijkp}(-t)=F_{jikp}(t)$. Following the integration over momentum, 
we derive that $F_{ij}(t)\approx \delta_{ij} F_{ii}(t)$, thereby leading to 
\begin{equation}
    \textrm{Im}\tilde{\chi}_{ij}(\omega)=\frac{\delta_{ij}}{2\hbar}[\tilde{F}_{ii}^{+-}(\omega)-\tilde{F}_{ii}^{-+}(\omega)].
\end{equation}
To clarify the relationship between $\tilde{F}_{ii}^{-+}({\omega})$ and $\tilde{F}_{ii}^{+-}(\omega)$, observe that 
\begin{align}
    \langle \hat{A}_{ikp}^\dagger(0)\hat{A}_{ikp}(t)\rangle&=\frac{1}{Z}\textrm{Tr}\left[e^{-\beta K} \hat{A}_{ikp}^\dagger(0)\hat{A}_{ikp}(t)\right]\nonumber\\
    &=e^{-\beta\Delta\mu}\langle \hat{A}_{ikp}(t-i\beta\hbar)\hat{A}_{ikp}^\dagger(0)\rangle,
\end{align}
where $K=H_{SC}+H_K-\sum_{k\sigma} \mu_\sigma c_{k\sigma}^\dagger c_{k\sigma}$ denotes the grand canonical Hamiltonian. 
In the final expression, we utilize the relation $e^{\beta K}\hat{A}_{ikp}(t)e^{-\beta K}=e^{-\beta\Delta\mu}\hat{A}_{ikp}(t-i\beta\hbar)$, 
subject to the condition $[H_{SC}+H_K,\sum_\sigma\mu_\sigma N_\sigma]=0$. An analogous methodology is applied to the correlations involving 
$\hat{B}_{ikp}(t)$ and $\hat{C}_{ikp}(t)$, yielding $F_{iikp}^{+-}(t)=e^{-\beta\Delta\mu}F_{iikp}^{-+}(t-i\beta\hbar)$. Ultimately, through 
the application of the Fourier transform, we attain 
\begin{equation}
    \tilde{F}_{ii}^{+-}(\omega)=e^{-\beta(\hbar\omega+\Delta\mu)}\tilde{F}_{ii}^{-+}(\omega),
\end{equation}
culminating in Eq. (\ref{eq.imchitilde}).

\section{Retarded Green's function}
\label{appendixB}
The imaginary time formalism offers a direct methodology to determine the retarded Green's function. In this context, 
the retarded Green's function is derived from the analytical continuation of $\tilde{\mathcal{G}}_q(i\nu_l)$, wherein the imaginary time 
Green's function is defined as
\begin{align}
    \hbar\mathcal{G}_r(\tau)&=-\langle T_\tau f_r(\tau)f_r^\dagger(0)\rangle_z\nonumber\\
    &=-\frac{1}{N_u}\sum_q\langle T_\tau f_q(\tau)f_q^\dagger(0)\rangle_z,
\end{align}
and $f_r(\tau)=e^{H_z\tau/\hbar}f_r e^{-H_z\tau/\hbar}$. $T_\tau$ represents the time-ordering operator for imaginary time, and the average
is evaluated with the full Hamiltonian $H_z=H_K+V_z$. Note that, employing the interaction picture, we obtain
\begin{equation}
    \hbar\mathcal{G}_r(\tau)=-\frac{\langle T_\tau\hat{f}_r(\tau)\hat{f}_r^\dagger(0)S(\beta)\rangle_0}{\langle S(\beta)\rangle_0},
\end{equation}
where the caret denotes the time evolution as dictated by $H_K$, and $S(\beta)=T_\tau\exp[-\int_0^{\beta\hbar}d\tau^\prime\hat{V}_z(\tau^\prime)/\hbar]$ 
represents the $S$ matrix associated with the perturbation $V_z=g_K (f_r^\dagger f_r-1/2)$. Note that the interaction term is confined to a 
single site, thus obviating any summation over the entire lattice, which would otherwise result in a trivial outcome. The magnitude of the potential is 
characterized by the coupling constant $g_K=-4J$, which is of the same order as the free Hamiltonian. Fortunately, we can determine the perturbed Green's 
function, including all orders of the potential term. Considering only the connected Feynman diagrams, Dyson's sum establishes that the perturbed Green's 
function is expressed in terms of the unperturbed Green's function as follows
\begin{align}
    \tilde{\mathcal{G}}_r(i\nu_l)&=\tilde{\mathcal{G}}_r^0(i\nu_l)+\tilde{\mathcal{G}}_r^0(i\nu_l)g_K\tilde{\mathcal{G}}_r^0(i\nu_l)+\ldots\nonumber\\  
    &=\frac{\tilde{\mathcal{G}}_r^0(i\nu_l)}{1-g_K\tilde{\mathcal{G}}_r^0(i\nu_l)}.
\end{align}
where $\tilde{\mathcal{G}}_r(i\nu_l)=N_u^{-1}\sum_q\tilde{\mathcal{G}}_q(i\nu_l)$. The unperturbed Green's function is readily obtained, being expressed by
\begin{align}
    \hbar\mathcal{G}_q^0(\tau)&=\cos^2\psi_qe^{-\nu_q\tau}[f(E_q)-\theta(\tau)]-\nonumber\\
    &-\sin^2\psi_q e^{\nu_q \tau}[f(E_q)-\theta(-\tau)],
\end{align}
where $f(E_q)=(e^{\beta E_q}+1)^{-1}$ denotes the Fermi-Dirac distribution applicable to complex fermions with energy $E_q=\hbar\nu_q$. 
In the imaginary frequency space, we derive 
\begin{equation}
    \tilde{\mathcal{G}}_q^0(i\nu_l)=\frac{\cos^2\psi_q}{i\hbar\nu_l-E_q}+\frac{\sin^2\psi_q}{i\hbar\nu_l+E_q},
\end{equation} 
with $\nu_l=(2l+1)\pi/\beta\hbar$, $l\in\mathbb{Z}$, representing the fermionic frequencies. Therefore, the 
analytical continuation $i\nu_l\to\nu+i\varepsilon$ yields the retarded Green's function 
\begin{equation}
    \tilde{G}_{\textrm{ret},q}^0(\nu)=\frac{\hbar\nu+E_q\cos2\psi_q+i\varepsilon}{(\hbar\nu+i\varepsilon)^2-E_q^2},
\end{equation}
where $E_q\cos 2\psi_q=2\textrm{Re}(S_q)$. Summing over the momentum results in Eq. (\ref{eq.G0retexact}).

\bibliographystyle{apsrev4-2}
\bibliography{bibliography.bib}
\end{document}